\documentclass{article}[12pt]

\usepackage{mathtools,amssymb,amsmath}
\usepackage{color,bm,xcolor}
\usepackage{multirow}
\usepackage{bm}
\usepackage{bbm}
\usepackage{indentfirst}
\usepackage{natbib}
\usepackage{subcaption}
\usepackage[title]{appendix}
\usepackage[colorlinks=true, allcolors=blue]{hyperref}
\usepackage{comment}
\DeclareMathOperator*{\argmin}{argmin}   

\newcommand{\bone}{\mathbbm{1}}

\newcommand{\tGamma}{\text{Gamma}}
\newcommand{\mA}{\mathcal{A}}
\newcommand{\mI}{\mathcal{I}}
\newcommand{\mM}{\mathcal{M}}
\newcommand{\mD}{\mathcal{D}}
\newcommand{\mAf}{\mA_{\text{final}}}

\newcommand{\hba}{\hat{\boldsymbol{a}}}
\newcommand{\ha}{\hat{a}}

\newcommand{\hatt}{\hat{t}}

\usepackage{mathtools}

\newcommand{\iid}{\stackrel{\mathclap{iid}}{\sim}}

{
	\title{\bf   SIMBA -- A Bayesian Decision Framework for the Identification of Optimal Biomarker Subgroups for   Cancer Basket Clinical  Trials }
	\author{
 	  Shijie Yuan\thanks{Department of Statistics and Data Sciences, The University of Texas at Austin, Austin, USA}\ , 
      Jiaxin Liu\thanks{BayeSoft Inc., Shanghai, CHN}\ ,  
      Zhihua Gong\thanks{LaNova Medicines Ltd., Shanghai, CHN}\ ,
      Xia Qin\footnotemark[3]\ , Crystal Qin\footnotemark[3]\ ,
      Yuan Ji\thanks{Department of Public Health Sciences, The University of Chicago, Chicago, USA; Corresponding email: koaeraser@gmail.com}, and Peter M\"uller\thanks{Department of Mathematics, The University of Texas at Austin, Austin, USA}\ 
 	}
	\date{\today}
}

\begin{document}
\maketitle

\begin{abstract}
We consider basket trials in which a biomarker-targeting drug may be efficacious for patients across different disease indications.  Patients are enrolled if  their cells exhibit some levels of biomarker expression. The threshold level is allowed to vary by indication. The proposed SIMBA method uses a decision framework to identify
optimal biomarker subgroups (OBS) defined by an optimal biomarker
threshold for  each  indication. The optimality is achieved through
minimizing a posterior expected loss that
 balances  estimation accuracy and investigator
 preference for  broadly effective therapeutics.
A Bayesian hierarchical model is proposed to  adaptively borrow information across  indications  and enhance the accuracy in the estimation of the OBS. The operating characteristics of SIMBA are assessed via simulations and compared against  a simplified version and an existing alternative method, both of which do not borrow information.  SIMBA is expected to improve  the  identification of  patient  sub-populations  that may benefit from a biomarker-driven therapeutics.    
\end{abstract}

\textbf{\textit{Keywords}}: Bayesian hierarchical model;   Cancer trials;  Biomarker threshold;   Precision oncology;  Subgroup identification.

\section{Introduction} \label{sec:intro}
In oncology, novel molecular or immune therapeutics are
potentially effective in an indication-agnostic fashion  when  their
underlying mechanism of action  targets tumor biomarkers that are
present across different cancers.
Basket trials are frequently used
 as efficient study designs to test
such  therapeutics in multiple indications simultaneously.
At the same time,
cancer patients are expected to exhibit heterogeneous response
to the therapeutics in a biomarker-dependent fashion. For example, a
biomarker positive subgroup may respond better to the treatment than a
biomarker negative subgroup.
These subgroups can be based on thresholding the expression
level of the related biomarker (e.g., a gene or protein), which is
reported as a continuous measurement.
For example,  many check-point inhibitors like Pembrolizumab are
approved for the treatment of adult and pediatric patients if their
PD-L1 expression is above a certain threshold
\citep{maio2022pembrolizumab,le2023pembrolizumab}. 
Therefore,  for novel therapeutics targeting a specific
biomarker  it is of interest to identify biomarker thresholds that define different patients subgroups with
clinically relevant differences in prognostic effects
\citep{samstein2019tumor,casarrubios2022tumor},
 and this is often needed in the context of basket trial designs. 

In recent years,
 several statistical methods have been proposed to  
address these issues
\citep{freidlin2005adaptive,jiang2007biomarker,zeileis2008model,lipkovich2011subgroup,scher2011adaptive,lu2015subgroup,thall2021adaptive,luo4167735adaptive,zhang2023robust}.
For example, \citet{freidlin2005adaptive} use logistic regression
models on biomarker expression levels, assuming a continuous gradient
in the log-odds of response rate with higher levels. They classified
patients as part of the positive subgroup if their predicted treatment
odds ratio exceeds a specified threshold.
\citet{jiang2007biomarker}  determine a  biomarker threshold for a sensitive subpopulation by maximizing the log-likelihood function over candidate cutoffs and  bootstrapping   confidence intervals.  
The Model-Based Recursive Partitioning (MOB) method implements partitioning by repeatedly testing for parameter instability across covariates using score-based fluctuation tests and splitting the data at covariate thresholds that maximize improvement in model fit \citep{zeileis2008model}.
The SIDES method \citep{lipkovich2011subgroup}, designed for randomized controlled trials, implements partitioning by recursively splitting the population based on covariates to identify subgroups with enhanced treatment effects, using test statistics to guide and validate each split. 
\citet{luo4167735adaptive} defined candidate subgroups based on
biomarker expression values (e.g., 25th, 50th, and 75th percentiles), identifying the optimal threshold by   minimizing a  log-rank test statistics for each candidate subgroup.
 In related problems 
Bayesian parametric and nonparametric models have  also  been proposed
for  early-phase clinical trials  \citep{berry2013bayesian,
  neuenschwander2016robust, zhou2021robot,lyu2023muce,chen2023ibis}
with a main focus on information borrowing  to better estimate
response rates or success probabilities across indications.    
 
Motivated by these  considerations and building on existing methods,  we propose a
\textbf{S}ubgroup 
\textbf{I}dentification \textbf{M}odel for \textbf{BA}sket trials
(SIMBA) based on a Bayesian decision framework, aiming to identify
optimal biological subgroups (OBS) across indications.  SIMBA extends
existing methods by
 setting up an inference approach that achieves several objectives
in  a single model-based decision framework.
First, SIMBA considers modeling indication-specific biomarker
thresholds to allow flexible prognostic compositions for different
indications. For example, it is possible that a therapeutic is
effective for all patients in one indication but only a subgroup
in another indication.
Second, SIMBA applies a Bayesian hierarchical model that adaptively
borrows information  
across  indications  to enhance the accuracy of threshold estimation and  to facilitate the identification of patient subgroups.
Third,   SIMBA employs a multi-stage decision framework, building upon
 methods proposed in  \citet{xu2020asied} and
\citet{lalonde2007model}, to  allow for   subgroup-specific go-no/go
decisions.
Central to this framework are the concepts of a Lower Reference Value
($LRV$) and a Target Value ($TV$)  as  desirable response rates.
$LRV$ represents a ``dignity line" for the development of a drug,
essentially setting the minimum threshold for proceeding with its
development, and  $TV$ provides a market-based estimate of the
compound's commercial viability, offering a target for its desired
efficacy.  In SIMBA, the response rates of identified positive
subgroups and the overall population (all-comers) are  estimated.
Should the response rates for both the positive subgroup and
all-comers fall below $LRV$ for an indication, further investigation
in that indication may not be  advised. 
Conversely, if the response rates exceed $LRV$, a  comparison with
$TV$ determines whether the treatment is recommendable
for an all-comers population or specifically for the positive
subgroup.   Importantly, the decision framework allows trial
investigators to  incoporate a preference for  finding a large
OBS and 
at the same time  to insist on  a high response rate for the
biomarker-positive subgroup. This feature provides added flexibility
and optimality in practical drug development. We will  validate 
SIMBA's properties using simulated and real-world examples.  

The remainder of the article is organized as follows. Section \ref{sec:proposed_design}   describes the proposed statistical framework while Section \ref{sec:trial_design} describes the SIMBA   method.   In Section \ref{sec:simluation}, we conduct simulation studies to demonstrate the operating characteristics of SIMBA, 
 followed by an application of the method to a real dataset from a cancer trial in Section \ref{sec:example}. 
We end the paper with a discussion in Section \ref{sec:discussion}.  
 
\section{Statistical Framework} \label{sec:proposed_design}

\subsection{Probability Model} \label{sec:prob_mod}
Consider a  basket  trial that evaluates a new drug in $I$ different
indications. Suppose $N_i$ patients   from indication $i$ are enrolled
to the trial, and biomarker
expression level $X_{ij}$ is measured for patient $j$ in indication
$i$, $i = 1, \cdots, I$, $j = 1, \cdots, N_i$.  
Assume that patients with $\{X_{ij} > x_i\}$ belong to the positive
subgroup with a response rate $p_{i+}$, and those with $\{X_{ij} \leq
x_i\}$ belong to the negative subgroup with a response rate $p_{i-}$.
Here, $x_i$ is considered a random and unknown threshold  for 
each indication $i$.  Without loss of generality, we assume $p_{i+} \ge
p_{i-}$ implying a positive association between the biomarker
expression and response rate.


 We consider a model-averaging framework allowing the two response rates to be either the same (i.e., no subgroups) or different (i.e., subgroups with different response rates). 
Let $M_i \in \{\mM_1,\mM_2\}$ denote two sub-models   for   indication $i$ where $p_{i+} = p_{i-} = p_i$ under  $M_i = \mM_1$  and $p_{i+} > p_{i-}$  under  $M_i = \mM_2$.   
Here, $p_i$  denotes   the response rate of all-comers in indication $i$ under $\mM_1$.  Conditional on $M_i$,  we assume the response of patient $j$ in indication $i$, $Y_{ij}$, follows a mixture of two Bernoulli distributions
\begin{equation}\label{eq:like}
    Y_{ij} \mid X_{ij},x_i,p_i,p_{i+},p_{i-},M_i \sim \bone(M_i = \mM_1) \cdot \text{Bern}(p_i) + \bone(M_i = \mM_2)\cdot\text{Bern}\Big( \bone(X_{ij}>x_i) \cdot p_{i+} + \bone(X_{ij} \leq x_i) \cdot p_{i-} \Big),
\end{equation}
where $\bone(\cdot)$ is an indicator function.

 We complete model specification with prior assumptions. 
We start with a Bernoulli prior for $M_i \in \{\mM_1, \mM_2\}$
as $p(M_i = \mM_1) = p_M$. 
A beta (hyper-)prior on $p_M$
\begin{equation}\label{eq:prior_pM}
    p_M \sim \text{Beta}(a,b)
\end{equation}
induces (marginal) dependence across $M_i$'s.   Next, let
$\theta_{i+} \equiv \text{logit}(p_{i+})$  and
$\theta_{i-} \equiv \text{logit}(p_{i-})$. 
Under sub-model $M_i = \mM_1$
let 
$\theta_i = \theta_{i+} = \theta_{i-} $, i.e.
$\theta_i \equiv \text{logit}(p_i).$ 
We assume a prior for $\theta_i$ given by  
\begin{align} \label{eq:prior_theta_m1}
    \theta_i \mid  M_i= \mM_1   \sim N(\mu_{\theta}, \sigma_{\theta}^2).
\end{align}
Under sub-model  $M_i = \mM_2$, noting that $p_{i+} > p_{i-}$   and therefore 
$\theta_{i+} > \theta_{i-}$, we  reparametrize   
$\theta_{i+}= \theta_{i-} + \delta_i,$  with $\delta_i > 0$  and  
\begin{align} \label{eq:prior_theta_m2}
    \begin{gathered}
        \theta_{i-} \mid  M_i = \mM_2   \sim N(\mu_{\theta -},\sigma_{\theta -}^2),\\
        \delta_i \mid  M_i=\mM_2  \sim \tGamma(a_{\delta},b_{\delta}).
    \end{gathered}
\end{align}
 Recall that biomarker thresholds $x_i$ are treated as unknown
parameters. We assume  an exchangeable prior  where conditional on
$\mu_x$ and $\sigma^2_x$ and under model $\mM_2$, $x_i$'s are
independent 
$$x_i \mid \mu_x,\sigma_x^2,  M_i = \mM_2  \sim N(\mu_x,\sigma_x^2)$$
with $\mu_x$ being a mean biomarker threshold.
We  complete the model with hyperpriors 
\begin{align} \label{eq:prior_x_m2}
    \begin{gathered}
        \mu_x \mid  \mM_2 \sim N(0,\sigma^2), \\
        \sigma_x \mid  \mM_2  \sim \text{Half-Cauchy}(0,\gamma).
    \end{gathered}
\end{align}
The use of the half Cauchy distribution prior follows
 a recommendation in 
\citet{Gelman2006}, for  
being weakly informative and computationally convenient.

\subsection{Posterior Inference}
Let 
$\boldsymbol{Y}$, 
$\boldsymbol{X}$, 
$\boldsymbol{M}$, $\boldsymbol{\theta}$, $\boldsymbol{\theta}_{-}$, $\boldsymbol{\delta}$, and $\boldsymbol{x}$ denote the set of all $Y_{ij}$'s, 
$X_{ij}$'s, 
$M_i$'s, 
$\theta_i$'s, 
$\theta_{i-}$'s, 
$\delta_i$'s, and 
$x_i$'s, respectively.
The joint posterior distribution of the parameters is given by
\begin{equation*}
    \begin{aligned}
        p(p_M, \boldsymbol{M}, \boldsymbol{\theta}, \boldsymbol{\theta}_{-}, \boldsymbol{\delta}, \boldsymbol{x}, \mu_x, \sigma_x \mid 
        \boldsymbol{Y}, \boldsymbol{X}) \propto  & \pi(\boldsymbol{M} \mid p_M)\pi(p_M) \times \prod_i\left[\prod_j f_1(Y_{ij} \mid \theta_i)\pi(\theta_i)\right]^{\bone(M_i = \mM_1)} \\
        &\prod_i\left[\prod_{j}f_2(Y_{ij} \mid   X_{ij}, x_i, \theta_{i-}, \delta_i)  \pi(\theta_{i-})  \pi(\delta_i)  \pi(x_i \mid \mu_x, \sigma_x)  \pi(\mu_x)  \pi(\sigma_x) \right]^{\bone(M_i=\mM_2)}
    \end{aligned}
\end{equation*}
where $\pi(\cdot)$ represents the corresponding prior densities  in Section \ref{sec:prob_mod},  
$$\quad f_1(Y_{ij} \mid    \theta_i) = \left[\frac{e^{\theta_i}}{1+e^{\theta_i}}\right]^{Y_{ij}}\left[\frac{1}{1+e^{\theta_i}}\right]^{1-Y_{ij}}, \text{ and} $$
{\small
$$f_2(Y_{ij} \mid  X_{ij}, x_i, \theta_{i-}, \delta_i) = 
\left\{\left[\frac{e^{\theta_{i-}}}{1+e^{\theta_{i-}}}\right]^{Y_{ij}}\left[\frac{1}{1+e^{\theta_{i-}}}\right]^{1-Y_{ij}}\right\}^{\bone(X_{ij} \leq x_i)}
\left\{\left[\frac{e^{\theta_{i-} + \delta_i}}{1+e^{\theta_{i-} + \delta_i}}\right]^{Y_{ij}}\left[\frac{1}{1+e^{\theta_{i-} + \delta_i}}\right]^{1-Y_{ij}}\right\}^{\bone(X_{ij} > x_i)}.
$$
}

 A posterior Monte Carlo sample of  the unknown parameters,
\begin{equation} \label{eq:MCMCsamples}
    \mathcal{S} \equiv \left\{\left(p_M^{(t)},\boldsymbol{M}^{(t)},  \boldsymbol{\theta}^{(t)}, \boldsymbol{\theta}_{-}^{(t)}, \boldsymbol{\delta}^{(t)}, \boldsymbol{x}^{(t)}, \mu_x^{(t)}, \sigma_x^{(t)}\right), \; t = 1, \cdots, T\right\},
\end{equation}
is obtained through Markov chain Monte Carlo (MCMC)  simulations,
where $T$ denotes the maximum number of MCMC iterations.  We
 use standard transition probabilities, including 
Gibbs and a Metropolis-Hastings hybrid algorithm to sample the
parameters from their conditional posterior distributions.   

\section{Proposed   SIMBA Method  } \label{sec:trial_design}
\subsection{ Decision Framework}\label{sec:dec_frame}

The objective of the basket trial is to explore the efficacy of the treatment   across a range of different disease indications.  For each indication, the treatment may be considered  efficacious for the entire patient population,  a biomarker subgroup, or neither.    In addition,  
at the end of the trial there may be insufficient evidence to reach a conclusion,   in which case the trial may be deemed inconclusive. To this end, we consider four  possible  decisions for each indication based on the observed data:  
``\textbf{S}top and do not recommend  the drug, ''
``\textbf{INC}onclusive,'' ``\textbf{R}ecommend for \textbf{A}ll
patients,'' and ``\textbf{R}ecommend for the \textbf{P}ositive
subgroup''.
Denote the
 action set of  four final decisions by  $\mAf \equiv
\{S, INC, RA, RP\}$.  

We  propose a decision framework to derive an optimal decision
$A_i^\star \in \mAf$ for each indication $i$.  Casting inference as a decision problem
\citep{berger2013statistical,robert2007bayesian} best reflects the nature of the problem as a decision maker chosing an action, rather than parameter estimation with generic estimation loss, and also makes it easy and more natural to set up the joint inference on $M_i$
and $(p_{i+},p_{i-},p_i)$. 
In the first step, following \cite{guo2017bayesian}, we obtain an optimal decision $\hba_i = (\ha_{i}, \ha_{i+}, \ha_{i-})$ for indication $i$ that determines the optimal sub-intervals   into   which the response parameters fall for all comers, positive subgroup (+), and negative subgroup (-), respectively. 
Second, we map $\hba_i$ to a single final decision $ A_i^\star \in
\mAf$ for indication $i$   based on a loss function that encompasses
three different components related to estimation accuracy and
investigator preferences.   
 We find it convenient to introduce both representations, $\hba_i$ and
$A_i^\star$ for the decision. For the loss function to be introduced
later it is easier to index the loss function by actions linked to the
reported intervals $\hba_i$. We then introduce the (deterministic)
mapping to $A_i^\star$ to link those to the reported actions on each indication
about stopping versus continuation.

To begin, we consider a decision framework similar to
\citet{guo2017bayesian} and derive the optimal decision
$\hba_i$. \citet{guo2017bayesian} proposed dividing the space $(0,1)$
for toxicity probabilities into sub-intervals with equal length in a
model-selection framework in order to ``blunt the Occam's razor", a
technique that avoids making biased dose-finding decisions.  Next we inspect whether  to decide whether the efficacy response rates $p_{i}$,
$p_{i+}$, or $p_{i-}$ are in $(0, LRV_i)$, $(LRV_i, TV_i)$, or $(TV_i,
1)$ as each interval implies a different decision. Specifically, the
interval $(0, LRV_i)$ implies that the response rate is too low to be
beneficial to patients, the interval $(TV_i, 1)$ implies that the
response rate is   sufficiently   high  to be beneficial, and the
interval $(LRV_i, TV_i)$ is in between.  Since the thresholds $LRV$
and $TV$ may differ across indications due to variations in disease
biology, patient characteristics, and response criteria, we use the
subscript $i$ to denote indication-specific parameters.  

A straightforward  next step  would  be to  identify one of the
three intervals  as  optimal if the posterior probability that the
efficacy rate of the drug falls into the interval is the
largest. However, because the lengths of the intervals are typically
different (LRV and TV are usually less than 0.5), such a framework
may   lead
to biased optimization
that prefers  shorter  intervals. As a remedy,  following \cite{guo2017bayesian}, we  partition the parameter space $(0,1)$  of
the response rates into sub-intervals of equal length. Without loss of
generality, we assume that $LRV_i$ and $TV_i$ are rounded to
 one decimal. 
Then, we set the sub-interval length to be the largest
common divisor (in the decimal sense) of $LRV_i$, $(TV_i - LRV_i) $,
and $(1 - TV_i) $.  
For example, if $LRV_i = 0.1$ and $TV_i = 0.3$, then the sub-interval length is $\epsilon_i = 0.1$. 
Let $K_i$ be the total number of sub-intervals in $(0,1)$  as a result of the partition.   Denote   $\{(c_0,c_1], (c_1,c_2], \cdots,  (c_{k_{i1} - 1},c_{k_{i1}}], \cdots, (c_{k_{i2} - 1},c_{k_{i2}}],\cdots,(c_{K_i - 1},c_{K_i})\}$ the partition, where $c_0 = 0$, $c_{k_{i1}} = LRV_i$, $c_{k_{i2}} = TV_i$, and $c_{K_i} = 1$.
Therefore, there are $k_{i1}$,   $(k_{i2} - k_{i1})$, and $(K_{i} - k_{i2})$   sub-intervals in the three intervals, $(0, LRV_i)$, $(LRV_i, TV_i)$, and $(TV_i,1)$, respectively.  Using the example above, the parameter space is divided into $K_i = 10$ intervals, $\{(0,0.1], (0.1,0.2], \cdots, (0.9,1)\}$, and we have $k_{i1} = 1$, $k_{i2} = 3$, and $K_{i} = 10$.

We  denote   $\mI \equiv \{1, 2, \ldots, K_i\} \times \{1, 2, \ldots, K_i\} \times \{1, 2, \ldots, K_i\}$  as  the action space of the  sub-interval indices $\boldsymbol{a}_i \equiv ( a_i,  a_{i+}, a_{i-} ) \in \mI$ for the all-comers, positive subgroup, and negative subgroup for indication $i$,   respectively. 
For example, $\{a_{i} = k\}$ means that the response rate $p_{i}$
for all comers   is  reported to be within   the $k$th
sub-interval $(c_{k-1}, c_k]$. 
Similarly,   $\{a_{i+} = k'\}$ indicates that the response rate
$p_{i+}$ for the positive subgroup is in $(c_{k'-1}, c_{k'}]$.   
Finally, by considering the action of  reporting $M_i$ to indicate one of 
the two potential models,  we denote $\boldsymbol{\phi}_i = (
p_{i},p_{i+},p_{i-},  M_i) \in [0,1]^3 \times \{\mM_1,\mM_2\}$ the
vector of  response rates and model indicators for indication $i$, and
define a loss function $l(\boldsymbol{a}_i, \boldsymbol{\phi}_i)$ as  
\begin{equation} \label{eq:loss}
l(\boldsymbol{a}_i,\boldsymbol{\phi}_i)\equiv \left\{
\begin{aligned}
   0,& &  \text{if }   p_{i} \in (c_{a_{i} - 1}, c_{a_{i}}] \text{ and } M_i = \mM_1;  \\
    0,& & \text{if } p_{i+} \in (c_{a_{i+} - 1}, c_{a_{i+}}]\text{, } p_{i-} \in (c_{a_{i-} - 1}, c_{a_{i-}}], \text{ and } M_i =  \mM_2;  \\
    1,& & \text{otherwise}.\\
\end{aligned}
\right.  
\end{equation}

In words, the loss is 0 if the action picks the right intervals   to   which the true response rates belong, under $M_i = \mM_1$ or $\mM_2$.   
The optimal interval indices $\hba_i$ for the three groups of indication $i$ are determined by minimizing the posterior expected loss,
\begin{equation}\label{aih}
\hba_i =  (\ha_{i}, \ha_{i+},\ha_{i-}) = \argmin_{\boldsymbol{a}_i \in
  \mI} \int l(\boldsymbol{a}_i, \boldsymbol{\phi}_i)
\pi(\boldsymbol{\phi}_i \mid  \mD)d\boldsymbol{\phi}_i.
\end{equation}
where $\mD$ denotes the observed data, $\mD = \{(X_{ij}, Y_{ij}); i = 1,\cdots, I, j = 1, \cdots, N_{i}\}$, and $\pi(\boldsymbol{\phi}_i \mid  \mD)$ is the posterior  distribution   of $\boldsymbol{\phi}_i$.
 Under the 0/1 loss function \eqref{eq:loss} the Bayes rule
\eqref{aih} reports the interval and model with the highest posterior
probability. 

Next, we propose a function $g(\hba_i): \mI \rightarrow \mAf$ that
maps the decision  $\ha_i$   to a trial action  $A_i^\star$ for each
indication $i$.  
The mapping $g(\hba_i)$ is defined in Table \ref{tab:map_tab} which
provides trial actions for all possible cases of $(\hat{a}_i,
\hat{a}_{i-}, \hat{a}_{i+})$. For example, if $\hat{a}_i \le k_{i1}$,
it means that the estimated $p_i$, the overall response rate, is less
than the LRV value for indication $i$.
If all three  indices $(\ha_i, \ha_{i-}, \ha_{i+})$  are $\le k_{i1}$, the response rate  of all patients for indication $i$  are believed to be lower than the LRV value, and the sensible decision is $S$, to stop and do not recommend the drug for the indication.   

\begin{table}[htbp]
  \centering
    \caption{Mapping function $g(\cdot)$ of optimal interval indices   $(\ha_{i}, \ha_{i+},\ha_{i-})$ to the    optimal trial  action $A^\star_i \in \mAf = \{S, INC, RA, RP\}$. Specifically, $\le k_{i1}$, $k_{i1} < \cdot \leq k_{i2}$, or $> k_{i2}$ means the corresponding response rate is less than the LRV value, between LRV and TV, or greater than the TV value for  indication $i$, respectively. 
     } \label{tab:map_tab}
    \begin{tabular}{l|c|c|c|c|c|c|c}
    \hline
    Decision & \multicolumn{3}{c|}{$\ha_{i-} \leq k_{i1}$} & \multicolumn{3}{c|}{$k_{i1} < \ha_{i-} \leq k_{i2}$} & \multicolumn{1}{c}{\multirow{2}[4]{*}{$\ha_{i-} > k_{i2}$}} \\
\cline{2-7}      $A_i^\star = g(\hba_i)$    & \multicolumn{1}{c|}{$\ha_{i} \leq k_{i1}$} & \multicolumn{1}{c|}{$k_{i1} < \ha_{i} \leq k_{i2}$} & $\ha_{i} > k_{i2}$  & \multicolumn{1}{c|}{$\ha_{i} \leq k_{i1}$} & \multicolumn{1}{c|}{$k_{i1} < \ha_{i} \leq k_{i2}$} & $\ha_{i} > k_{i2}$  &  \\
    \hline
    $\ha_{i+} \leq k_{i1}$ & \multicolumn{3}{c|}{$S$} & \multicolumn{4}{c}{   n/a   } 
    \\
    \hline
    $k_{i1} < \ha_{i+} \leq k_{i2}$ & \multicolumn{1}{c|}{$S$} & \multicolumn{2}{c|}{$INC$} & \multicolumn{3}{c|}{$INC$} &     n/a   \\
    \hline
    $\ha_{i+} > k_{i2}$ & \multicolumn{2}{c|}{$RP$} & $RA$    & \multicolumn{2}{c|}{$RP$} & $RA$    & $RA$ \\
    \hline
    \end{tabular}
\end{table}

Due the the constraints that $p_{i+} > p_{i-}$, not all combinations for the three optimal interval indices $(\ha_{i}, \ha_{i+},\ha_{i-})$ are possible. For example, it is impossible to have $\ha_{i+} \leq k_{i1}$ and $\ha_{i-} > k_{i2}$ as this would imply $p_{i+} < p_{i-}$,   and therefore the corresponding trial action is not available ( n/a ) as shown in Table \ref{tab:map_tab}.  

\subsection{Optimal Biomarker Subgroup} \label{sec:OBS} A trial action
$RP$ indicates that  the drug   is recommended  for patients  in
indication $i$ with positive biomarker expression levels, i.e. $X_{ij}
> x_i$. However, $x_i$ is
 another unknown parameter in our model. To carry out the
study design we therefore need to still chose a biomarker threshold
$t_i$. The choice of $t_i$ is informed by available information on
$x_i$, but in contrast to inference on $x_i$, the choice of $t_i$ is
framed as a decision problem. The main motivation for doing so is the
opportunity to explicitly consider related investigator preferences,
which are not well represented by commonly used estimation losses like
squared error loss.
We set up another decision problem with the following
composite loss function $l(t_i,x_i, \mD)$, which is chosen to reflect
the competing considerations and preferences: 
\begin{equation} \label{eq:loss_t}
    l(t_i,x_i, \mD) = \underbrace{1 - e^{-|t_i - x_i|}}_{l_1(t_i, x_i)} + \underbrace{w_1 \times \frac{1}{N_i} \sum_{j}\bone(X_{ij}\leq t_i)}_{w_1 \times l_2(t_i, \mD)} + \underbrace{ w_2 \times  \left[1 - \frac{\tilde{p}_{i+}(t_i)}{TV_i} \right] \cdot \bone\left(\tilde{p}_{i+}(t_i) < TV_i\right) }_{w_2 \times l_3(t_i, \mD)}.
\end{equation}
where $\tilde{p}_{i+}(t_i) = \frac{\sum_{j}\bone(X_{ij}> t_i) \cdot
  Y_{ij}}{\sum_{j}\bone(X_{ij}> t_i) }$ represents the empirical
response rate of the positive subgroup given the threshold $t_i$. The
first term $l_1(t_i,x_i)$ in \eqref{eq:loss_t} quantifies the
discrepancy between the estimated and true threshold. The second
term $l_2(t_i, \mD)$ imposes a penalty based on the sample size of the
negative subgroup, thereby encouraging more patients to be assigned to
the positive subgroup. 
The third term $l_3(t_i, \mD)$ penalizes deviations of the positive subgroup's response rate from the target value $TV_i$, ensuring that the estimated positive subgroup maintains a sufficiently high response rate. 
The weights  $w_1>0$   and  $w_2>0$   control the relative influence of the penalties in $l_2$ and $l_3$  to $l_1$ which is the penalty for estimation error. 

Therefore, the optimal biomarker threshold for the positive subgroup is estimated by minimizing the posterior expected loss,
\begin{equation} \label{eq:min_loss_t}
    \hatt_i =  \argmin_{t_i} \int l(t_i, x_i, \mD) \pi(x_i \mid  \mD)dx_i.
\end{equation}
  And the corresponding optimal biomarker subgroup, OBS  $=\{i: X_{ij} > \hat{t}_i\}$ for indication $i$. Investigators may consider using the OBS as the target patient population for subsequent drug development. 

The proposed  decision-theoretic framework to determine the  OBS has several advantages.
First,  as already mentioned, 
the explicit representation of investigator preferences for finding a large
group for efficacious biomarkers. Second, the setup avoids massive
multiple comparison correctoins that would arise in an estimation and
multiple testing framework. 

\subsection{Interim Analysis}
The proposed SIMBA framework
 was introduced assuming use for the final data analysis.
Another use case is  to apply SIMBA as an
interim analysis, with slightly different actions. To start, let $N_i$
denote the maximum sample size for each indication $i$ and assume an
interim analysis is conducted when $n_{i1} < N_i$ patients have been
enrolled with indication $i$. Let $n_{i2} = N_i - n_{i1}$.

The   trial  actions   for   the interim is slightly different including either stopping the trial or continue to enroll patients. Therefore, we consider 
``\textbf{S}top the trial,'' ``\textbf{E}nroll \textbf{A}ll-comers
(EA),'' and ``\textbf{E}nroll \textbf{P}ositive subgroup (EP)'', as
the actions in the interim.
 The last two actions imply continuation. We denote the full
action set as 
$  \mathcal{A}^{(R)}  = \{S, EA, EP\}$.  
Correspondingly, there is not the action $INC$ for the all-comers,
positive subgroup, and negative subgroup at the interim analysis.
Hereinafter, we use the superscript   $(R)$  to represent the interim analysis. 

Let   $\mD^{(R)} = \{(X_{ij}, Y_{ij}) \mid i = 1,\cdots, I, j = 1, \cdots, n_{i1}\}$  denote the observed data at the interim analysis. 
The optimal sub-interval indices at the interim analysis are determined as $\hba_i^{(R)}   = \argmin_{\boldsymbol{a}_i \in \mI} \int l(\boldsymbol{a}_i, \boldsymbol{\phi}_i) \pi(\boldsymbol{\phi}_i \mid  \mD^{(R)})d\boldsymbol{\phi}_i$. 
The   interim trial   optimal action $A_i^{\star(R)}$ for indication $i$ is determined by
\begin{equation} \label{eq:mapping_interim}
    A_i^{\star(R)}  = g^{(R)} \left( \hba_i^{(R)}\right) = \left\{
\begin{aligned}
    EA,& & \text{if } \ha_{i+}^{(R)} > k_{i1} \text{ and }\ha_{i-}^{(R)} > k_{i1},\\
    EP,& & \text{if } \ha_{i+}^{(R)} > k_{i1}, \ha_{i-}^{(R)} \leq k_{i1}, \text{ and } \ha_{i}^{(R)} > k_{i1}\\
    S,& & \text{if } \ha_{i+}^{(R)} > k_{i1}, \ha_{i-}^{(R)} \leq k_{i1}, \text{ and } \ha_{i}^{(R)} \leq k_{i1}\\
    S,& & \text{if } \ha_{i+}^{(R)} \leq k_{i1} \text{ and }\ha_{i-}^{(R)} \leq k_{i1}\\
\end{aligned}
\right.
\end{equation}
The mapping function \eqref{eq:mapping_interim} is also illustrated 
in Table \ref{tab:map_tab_interim}.   Since at interim, the focus is on either stopping the trial due to futility or continue otherwise, the mapping function $g^{(R)} $ is simpler. Specifically, if a response rate is deemed below the LRV value, the drug may be stopped for the corresponding patient population. That is the main rationale for the new mapping function in \eqref{eq:mapping_interim} and Table \ref{tab:map_tab_interim}.  
\begin{table}[htbp]
  \centering
  \caption{Mapping function $g^{(R)}(\cdot)$ of optimal interval indices to the overall optimal action in the interim analysis.} \label{tab:map_tab_interim}
    \begin{tabular}{l|c|c|c}
    \hline
    Decision & \multicolumn{2}{c|}{$\ha_{i-} \leq k_{i1}$} & \multicolumn{1}{c}{\multirow{2}[4]{*}{$\ha_{i-} > k_{i1}$}} \\
\cline{2-3}    $A_i^{\star(R)}  = g^{(R)}$     & $\ha_{i} \leq k_{i1}$     & $\ha_{i} > k_{i1}$     &  \\
    \hline
    $\ha_{i+} \leq k_{i1}$     & \multicolumn{2}{c|}{$S$} &  n/a  \\
\cline{1-4}    $\ha_{i+} > k_{i1}$     & $S$ & $EP$ & $EA$ \\
    \hline
    \end{tabular}
\end{table}
 As previously noted, not all combinations for the three optimal
interval indices $(\ha_{i}, \ha_{i+},\ha_{i-})$ are  possible   due to
the constraint $p_{i+} > p_{i-}$, and  impossible   combinations are
marked as  n/a   as shown in Table \ref{tab:map_tab_interim}.  


\subsection{Overall Design}
  The proposed SIMBA approach can be used a design for a clinical trial in the  following three steps.  
\begin{enumerate}
    \item Enroll patients in parallel to different indications and record their responses and biomarker profiles.
    \item When the sample size of indication $i$ reaches $n_{i1}$,   apply  the interim analysis for all the indications and obtain the optimal decisions $A_i^{\star(R)} \in \mA^{(R)}$. 
    \item At the end of the trial, apply the final analysis and obtain the optimal decision $A_i^\star \in \mAf$  and report the OBS.  
\end{enumerate}
Step 2 may be simplified if there are many indications in the trial. For example, one may set up just one interim analysis when half of the total sample size is enrolled across all the indications. 

\section{Simulation Study} \label{sec:simluation}

\subsection{Simulation Setup}\label{sec:simsetup}
 We carried out simulations  to evaluate the operating
characteristic of the proposed SIMBA approach based on 1,000
independently simulated trials.  We consider the simulated trial with
$I = 3$ indications, where a maximum of 50 patients is to be enrolled
per  indication $(N_i = 50),$    $i = 1,2,3$.  An interim analysis is
planned after $n_{i1} = 40$ patients, with an additional $n_{i2} = 10$
patients   to be enrolled if the indication is not stopped after
interim. For simplicity, we assume   that the enrollment rates across
all three indications are the same, and  the interim analysis will be
conducted simultaneously for all indications. 
The biomarker expression levels of patients $X_{ij}$ are assumed to
follow a standard normal distribution, i.e., $X_{ij} \sim N(0,1)$. The
true biomarker thresholds  for the three indications $\boldsymbol{x} =
(x_1,x_2,x_3)$ are set as $(-0.1, 0, 0.1)$. The weights $w_1$ and
$w_2$ in the loss function \eqref{eq:loss_t} are specified as 0.2 and
0.5, respectively.  

\paragraph{Simulation Scenario} 
For each indication, we assume $LRV_i = 0.1$ and $TV_i = 0.3$, $i =
1,2,3$, leading to $k_{i1} = 1$ and $k_{i2} = 3$.   We then consider
six scenarios, each specifying true response rates of   the  positive
$(p_{i+})$ and negative $(p_{i-})$ subgroups,   as well as the average
response rate for all patients  computed as
$p_{i} = \Pr(X_{ij} \le x_i) p_{i-} +   \Pr(X_{ij} > x_i) p_{i+}$.
Table \ref{tab:sc_optimal_decision}  displays the true response
rates of the positive and negative subgroups under each indication and
scenario. In the simulated trial, after a patient is enrolled,
their biomarker expression level $X_{ij}$ is  generated   and compared
to the true biomarker threshold $x_i$. If $X_{ij} > x_i$, i.e. the
patient is from the positive subgroup, their  binary efficacy
response is generated from a Bernoulli distribution with success
probability of $p_{i+}$; otherwise, the response is drawn from a
Bernoulli distribution with success probability $p_{i-}$.  

Table \ref{tab:sc_optimal_decision}  shows the optimal   trial
decision defined by Table \ref{tab:map_tab} applied to 
the decisions \eqref{aih} for each indication, which in turn are
based on the loss  function 
\eqref{eq:loss} substituting the true response rates instead of
posterior integration. 

For example, in scenario 3, the response rates for indication 1 are
$p_{1+} = 0.4$ and $p_{1-} = 0.1$. The overall response rate is then
calculated as $p_{1} = [1 - \Phi(-0.1)]  0.4 + \Phi(-0.1)  0.1 \approx
0.26$, where $\Phi(\cdot)$ represents the cumulative distribution function (CDF) of the standard normal distribution.  
Plugging the true value of $p_1$, $p_{1+}$, and $p_{1-}$ into the loss function \eqref{eq:loss}, we have the following:  under sub-model $\mM_1$, the optimal interval index for all-comers is $a_1 = 3$; under sub-model $\mM_2$, the optimal interval indices for positive and negative subgroups are  $a_{1+} = 4$ and $a_{1-} = 1$, respectively. 
Referring to Table \ref{tab:map_tab}, the optimal decision of indication 1 is $RP$,
recommending the drug for the biomarker positive subgroup.

\begin{table}[htbp]
  \centering
  \caption{True response rates for the positive and negative subgroups, along with the average response rate of all-comers, across the three indications. The corresponding final and optimal decisions for each of the six scenarios are provided, given the true thresholds of expression levels $(x_1,x_2,x_3) = (-0.1, 0, 0.1)$.}
    \begin{tabular}{c|ccc|ccc}
    \hline
    \multirow{2}[1]{*}{Scenario} & \multicolumn{3}{c|}{\{$p_{i-},p_{i},p_{i+}$\}}& \multicolumn{3}{c}{Optimal Decision} \\
\cline{2-7}          & Indication 1 & Indication 2 & Indication 3& Indication 1 & Indication 2 & Indication 3 \\
    \hline
    1     & \{0.05, 0.05, 0.05\} & \{0.05, 0.05, 0.05\} & \{0.05, 0.05, 0.05\} & $S$ & $S$ & $S$\\
    2     & \{0.2, 0.2, 0.2\} & \{0.2, 0.2, 0.2\} & \{0.2, 0.2, 0.2\} & $INC$ & $INC$ & $INC$\\
    3     & \{0.1, 0.26, 0.4\} & \{0.1, 0.25, 0.4\} & \{0.1, 0.24, 0.4\} & $RP$ & $RP$ & $RP$\\
    4     & \{0.1, 0.26, 0.4\} & \{0.1, 0.25, 0.4\} & \{0.1, 0.19, 0.3\} & $RP$ & $RP$ & $INC$\\
    5     & \{0.4, 0.4, 0.4\} & \{0.4, 0.4, 0.4\} & \{0.1, 0.24, 0.4\} & $RA$ & $RA$ & $RP$\\
    6     & \{0.2, 0.2, 0.2\} & \{0.1, 0.25, 0.4\} & \{0.1, 0.24, 0.4\} & $INC$ & $RP$ & $RP$\\
    \hline
    \end{tabular}
  \label{tab:sc_optimal_decision}
\end{table}

\paragraph{Model Parameters} 
We apply SIMBA under the following settings:
$a = b =1$ for the prior of $p_M$,
$\mu_{\theta} = -2.3$ and $\sigma_{\theta}^2 = 10^2$ in sub-model $\mM_1$, and $\mu_{\theta-} = -2.3$, $\sigma_{\theta-}^2 = 2^2$, $a_{\delta} = 30$, $b_{\delta} = 21.5$, $\sigma^2 = 2^2$, and $\gamma = 2.5$ in sub-model $\mM_2$. 
The  implied 
prior distributions of $p_i$ in sub-model $\mM_1$ and $p_{i-}$ and
$p_{i+}$ in sub-model $\mM_2$ are
 vague priors with low density across the unit interval and high
prior variances. The densities are shown in   Figure
\ref{fig:prior_p}.

 Next we  examine the impact of the prior setting, especially
 the implied borrowing of strength in the prior model
\eqref{eq:prior_x_m2}. 
We consider a simplified model in which prior
\eqref{eq:prior_x_m2}  is replaced by
assuming $x_i \iid N(\mu_x, \sigma_x^2)$  with fixed $\mu_x = 0$ and
$\sigma_x^2 = 3^2$.
We call the  simplified  model nb-SIMBA indicating
no-borrowing since the independent construction prevents borrowing
information across indications when estimating the biomarker
thresholds $x_i$'s.  The two  prior distributions of $x_i$ in SIMBA
and nb-SIMBA are compared in Figures \ref{fig:prior_x} and
\ref{fig:joint_prior_x}. 
 The figures highlight  that $x_i$
are dependent under SIMBA but independent under nb-SIMBA.

\subsection{Simulation Results}

\subsubsection{ Comparison of SIMBA and nb-SIMBA}
\begin{figure}[!htb]
    \centering
    \begin{subfigure}{.26\textwidth}
    \centering
    \includegraphics[width=\linewidth]{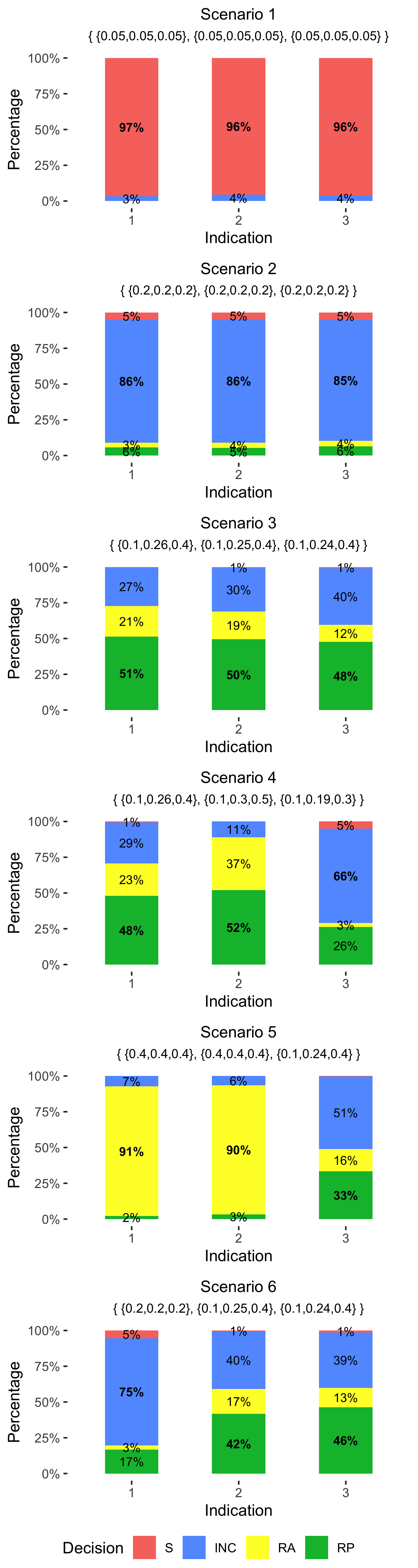}
    \caption{SIMBA}
    \end{subfigure}
    \begin{subfigure}{.26\textwidth}
    \centering
    \includegraphics[width=\linewidth]{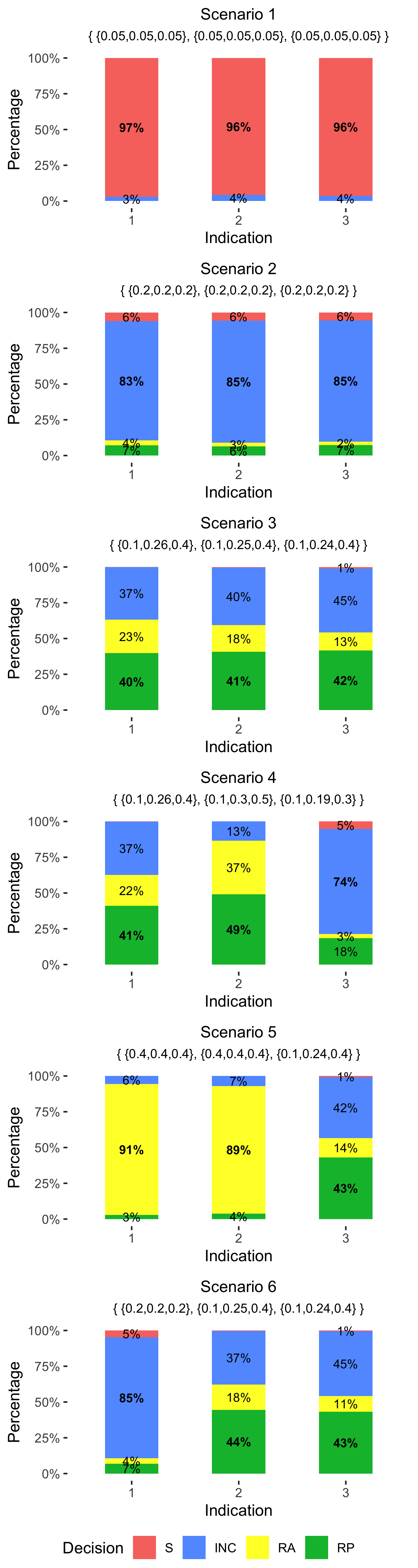} 
    \caption{nb-SIMBA}
    \end{subfigure}
    \begin{subfigure}{.39\textwidth}
    \centering
    \includegraphics[width=\linewidth]{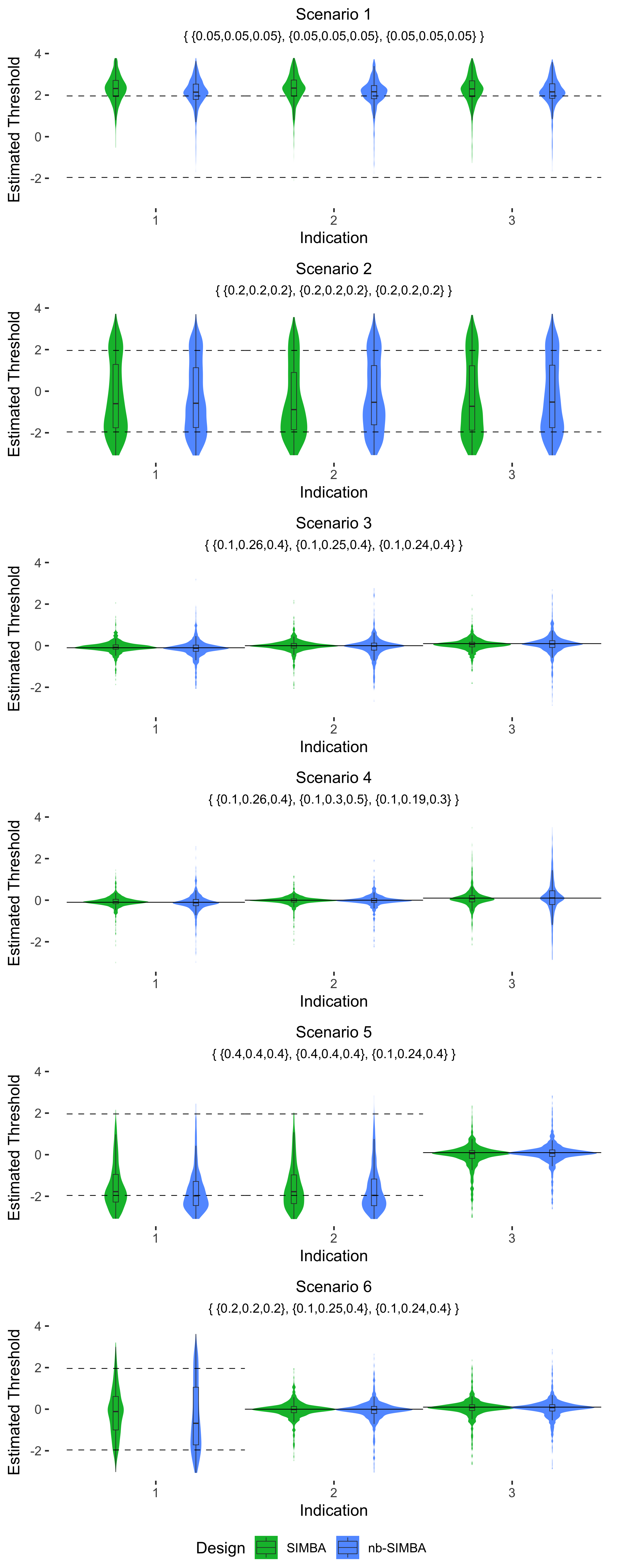} 
    \caption{Estimated Threshold}
    \label{fig:x_violin_xtrue2}
    \end{subfigure}
    \caption{Operating characteristics with the true thresholds of expression levels $(x_1,x_2,x_3) = (-0.1, 0, 0.1)$. The first two  column panels (a) and (b)  present  the percentages of four distinct final decisions under each scenario for SIMBA and nb-SIMBA, where the bold font denotes the optimal decision. The last panel (c) presents the violin   plots of the estimated $\hat{t}_i$  for SIMBA and nb-SIMBA. The dashed lines represent the 95\% range of simulated $X_{ij}$  values,   while the solid lines indicate the true biomarker threshold.}
    \label{fig:op_01}
\end{figure}

Operating characteristics
 for repeat simulation under scenarios 1 -- 6 using 
true thresholds of expression
levels $(x_1,x_2,x_3) = (-0.1, 0, 0.1)$ are shown in Figure
\ref{fig:op_01}.
 By way of a sensitivity analysis simulations with alternative
simulation truths for $(x_1,x_2,x_3)$ are shown Figures \ref{fig:op_0}
and \ref{fig:op_05}. 
We report the percentages  of repeat simulations  that report
the correct decisions at the end of trial and  show  violin
plots  of estimated biomarker thresholds $\hatt_i$ across 1,000
simulations.
Here, ``correct decision'' refers to the optimal
decisions reported in Table \ref{tab:sc_optimal_decision}.

In scenarios 1 and 2, both SIMBA and nb-SIMBA exhibit similar performance. In scenario 1, the optimal decision $S$ is selected in approximately 96\% of the simulations, while in scenario 2, the optimal decision $INC$ is chosen in approximately 85\% of the simulations. No subgroups exist in either scenario. As shown in Figure \ref{fig:x_violin_xtrue2}, the estimated biomarker thresholds are mostly near the upper or lower bounds of the simulated biomarker expression values, indicating that  most patients are likely classified into  a single group, consistent with true setting of both scenarios.   

Scenario 3, characterized by  higher response rates for the positive
subgroups, favors the decision $RP$ for all the three
indications. SIMBA  makes   this decision 51\%, 50\%, and 48\%  of the
times  across the three indications, respectively. In contrast,
nb-SIMBA  chooses  $RP$ 40\%, 41\%, and 42\%  of the times, which
suggests that SIMBA is more effective in recognizing the optimal
decision in this scenario.  

In scenario 4, although
 the simulation truth includes  
positive subgroups with higher response rates, the optimal decision of
indication 3 is $INC$.  
SIMBA chooses  $RP$ more frequently compared to nb-SIMBA in
indications 1 and 2, while SIMBA shows  a lower percentage of
selecting the optimal decision in indication 3. Furthermore, as shown
in Figure \ref{fig:x_violin_xtrue2}, the estimated biomarker
thresholds for SIMBA in scenarios 3 and 4 are closer to the true
biomarker threshold, indicating a higher probability of selecting the
correct threshold. 

In scenario 5,  the optimal decisions for indications 1 and 2 are $RA$
and $RP$ for indication 3.   SIMBA and nb-SIMBA exhibit similar
performance for indications 1 and 2.
However, because only one indication includes a positive subgroup with
a higher response rate, borrowing information across indications
 is not favorable here, and 
consequently SIMBA selects the optimal decision for
indication 3 at a lower rate. In scenario 6, the optimal decision is
$INC$ for indication 1, while it is $RP$ for indications 2 and
3. Although SIMBA more frequently selects the optimal decision for
indications 2 and 3, it is less likely to do so for indication 1,
where no subgroups exist. 

In scenarios 5 and 6, SIMBA and nb-SIMBA produce similar estimates of the biomarker threshold. 
SIMBA is more likely to select a threshold closer to 0 for indication 1 in scenario 6, presumably due to borrowing information from the other two indications.

Overall, SIMBA  performs favorably  across
scenarios where the positive subgroups of all three indications
exhibit substantially higher response rates. Even in scenarios where
some indications have the same response rate across all-comers, SIMBA
performs comparably to nb-SIMBA. 

\clearpage\newpage
\subsubsection{ Comparison  with   MOB}

We compare  SIMBA with  
the MOB method \citep{zeileis2008model}, implemented using the \texttt{partykit} package in R.  Recall the brief summary of MOB in Section \ref{sec:intro}. 
Logistic regression models are fitted within each node using the \texttt{mob()} function, allowing partitioning based on biomarker expression values, with an intercept-only logistic model fitted within each subgroup. Control settings are specified using \texttt{mob\_control()}, with the maximum tree depth set to 2 to restrict tree complexity for comparison purposes. Under this setting, a maximum of two subgroups can be formed for each indication.

In particular, MOB focuses on reporting presence or absence of biomarker subgroups under each indication.
There is no notion of recomemnding final decisions like $A_i^\star$ under SIMBA. Matching the inference target of MOB as the identification of relevant biomarker subgroups we reduce inference under SIMBA (and nb-SIMBA) to focus only on the identification of biomarker subgroups, which we implement as an indicator for $\Pr(M_i=\mM_2 \mid \mD) > \lambda$. In this way we can compare the inference under MOB and SIMBA, ignoring the more detailed reports under SIMBA on $A_i^\star$. 
 We select the thresholds $\lambda = 0.98$ and $\lambda_{\text{nb}} = 0.97$ such that the percentages of identifying the existence of subgroups for indication 1 in scenario 6 under SIMBA and nb-SIMBA match that under MOB, as indication 1 in scenario 6 represents the ``worst-case" scenario for SIMBA. 

\begin{figure}[!htb]
    \centering
    \begin{subfigure}{.42\textwidth}
    \centering
    \includegraphics[width=\linewidth]{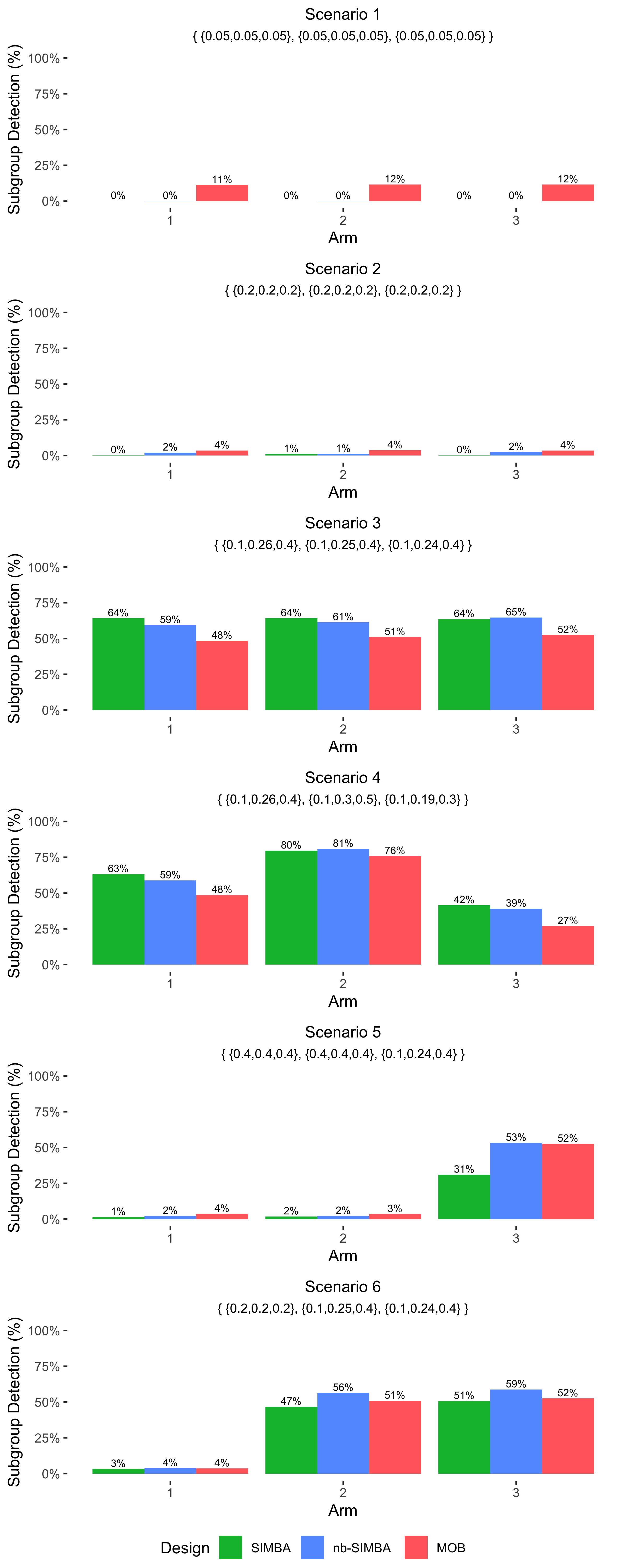}
    \caption{Percentage of Subgroup Identification}\label{fig:MOB_M2_prop}
    \end{subfigure}
    \begin{subfigure}{.42\textwidth}
    \centering
    \includegraphics[width=\linewidth]{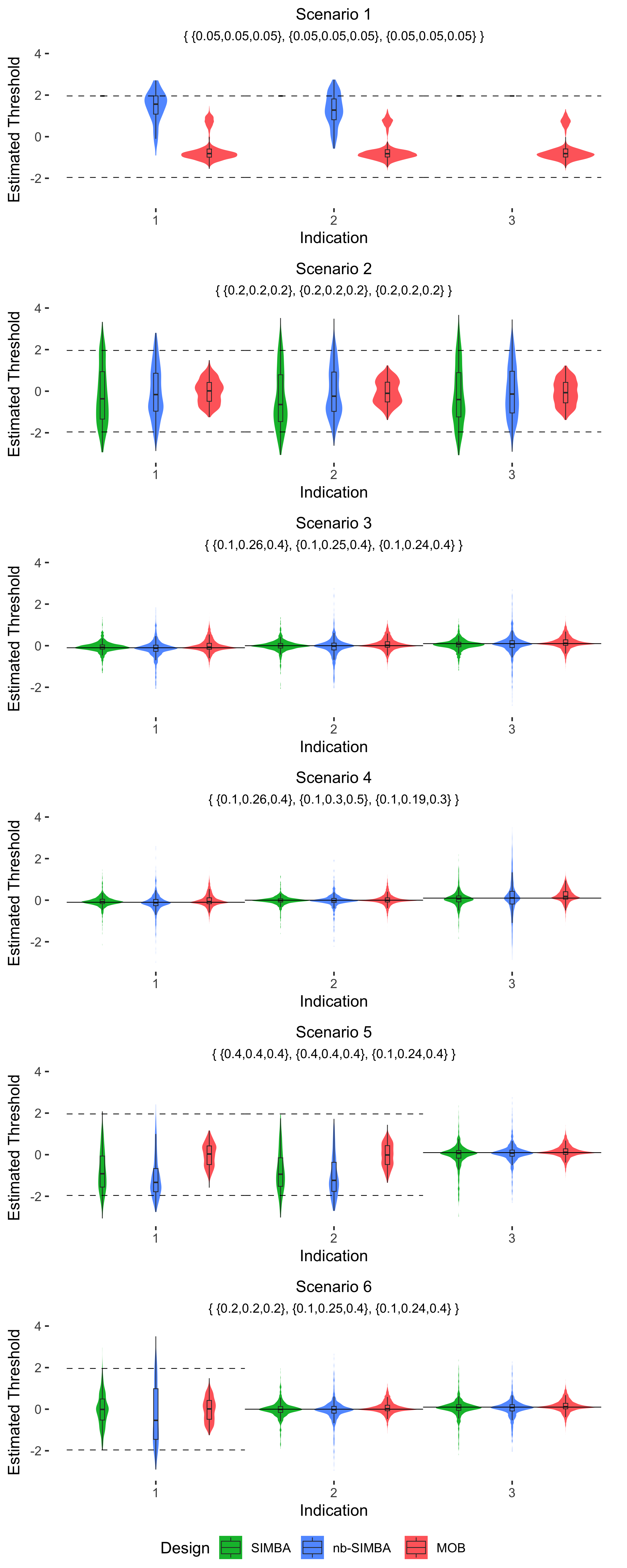} 
    \caption{Estimated Threshold with Subgroups Identified}\label{fig:MOB_x}
    \end{subfigure}
    \caption{ Operating characteristics with the true thresholds of expression levels $(x_1,x_2,x_3) = (-0.1, 0, 0.1)$. The first panel (a)  presents  the percentages of identifying the existence of subgroups within each indication across different scenarios. The panel (b) displays violin plots of the estimated $\hat{t}_i$ for  simulated trials that identified the existence of  subgroups. The dashed lines represent the 95\% range of simulated $X_{ij}$ values, while the solid lines indicate the true biomarker thresholds.} \label{fig:MOB_comp}
\end{figure}

Simulation results are shown in Figure \ref{fig:MOB_comp}. 
In scenario 1, both SIMBA and nb-SIMBA perform favorably compared to MOB,  as the correct decision is reporting no subgroups.  
A similar pattern is observed in scenario 2.  And even when  occasionally SIMBA identifies subgroups, the estimated $\hatt_i$ values are concentrated to the lower bound of the simulated biomarker expression values,  indicating that  most patients are classified into a single group, consistent with true setting of both scenarios. 

In scenarios 3 and 4, SIMBA identifies the  correct  subgroups with a higher frequency than MOB, while the estimated thresholds remain comparable between methods.

In scenario 5, because only one indication includes a
positive subgroup with a higher response rate, borrowing information across indications  may not benefit  here. As a result, SIMBA identifies  the correct  subgroups for indication 3 at a lower rate. Furthermore,  the estimated $\hatt_i$ values  under SIMBA  are again close to the lower bound of the simulated biomarker expression values. In scenario 6, the percentages of subgroup identification and the estimated thresholds for SIMBA and MOB are comparable.

\clearpage\newpage
\subsection{Sensitivity Analysis}
We conduct  some sensitivity analyses.
Here we focus on  different prior specifications for $\sigma_x$ in
\eqref{eq:prior_x_m2},
specifically the value of the hyperparameter $\gamma$,  since
$\sigma_x$ is known to affect the information borrowing across $x_i$'s
under the proposed Bayesian inference.     
 More sensitivity analyses with respect to the weights $w_1$ and $w_2$ in the loss function
\eqref{eq:loss_t} are shown in the appendix, in Figure
\ref{fig:w_compare}. 

We examine two  hyperparameter settings for $\gamma$ of the
half-Cauchy prior in \eqref{eq:prior_x_m2}, setting $\gamma = 1$ or
$\gamma = 4$, while keeping all other parameters  as before. 
Setting $\gamma = 1$ imposes stronger borrowing on the biomarker
threshold across indications compared to $\gamma = 2.5$, while $\gamma
= 4$ imposes weaker borrowing.  Additionally, we consider an inverse
gamma (IG) prior distribution for $\sigma_x$, assuming $\sigma_x \mid
\mM_2 \sim \text{ IG }(1, 1).$

\begin{figure}[!htb]
    \centering
    \begin{subfigure}{.48\textwidth}
    \centering
    \includegraphics[width=\linewidth]{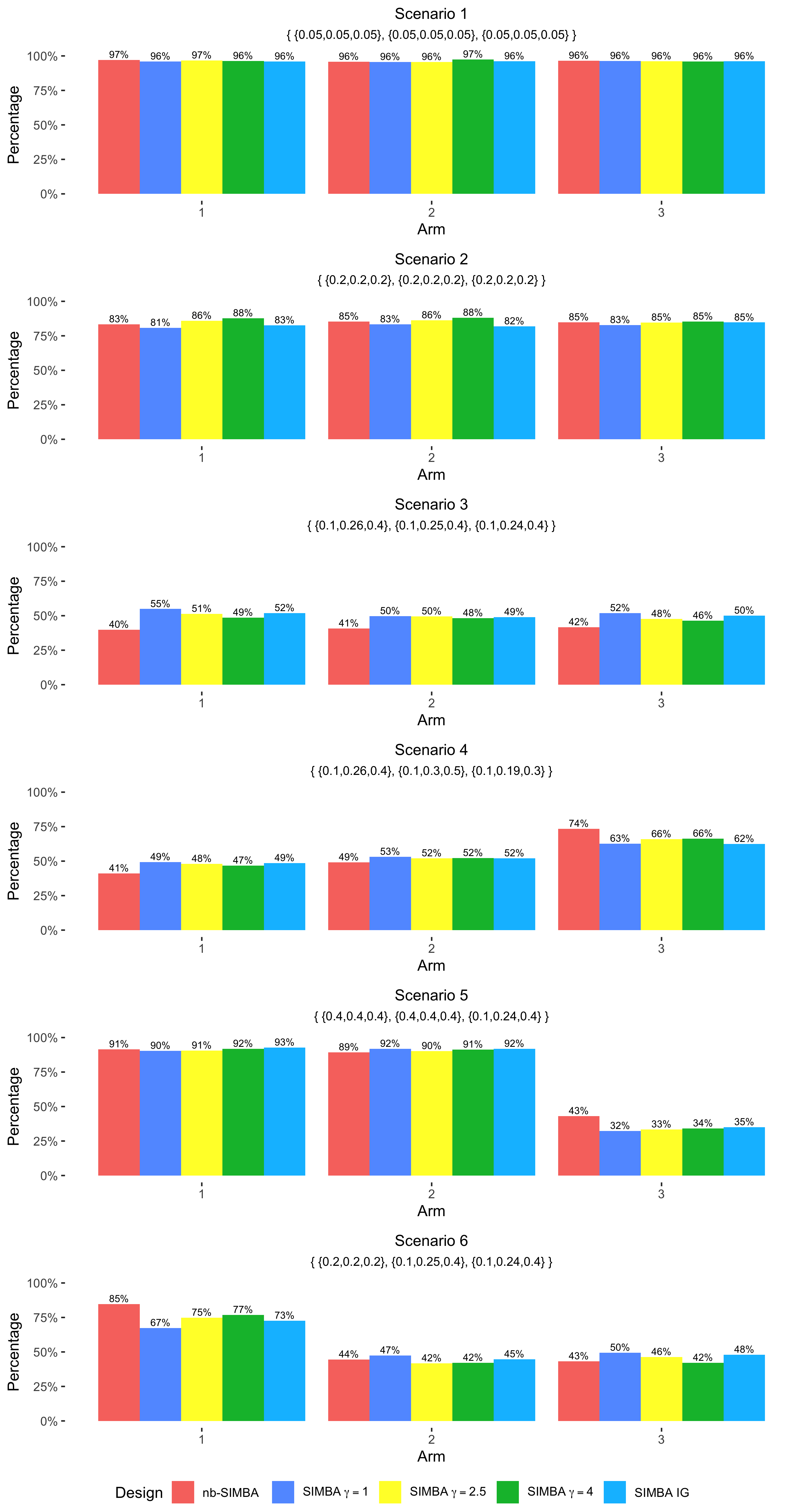}
    \caption{Percentages of choosing the optimal decision}
    \end{subfigure}
    \begin{subfigure}{.48\textwidth}
    \centering
    \includegraphics[width=\linewidth]{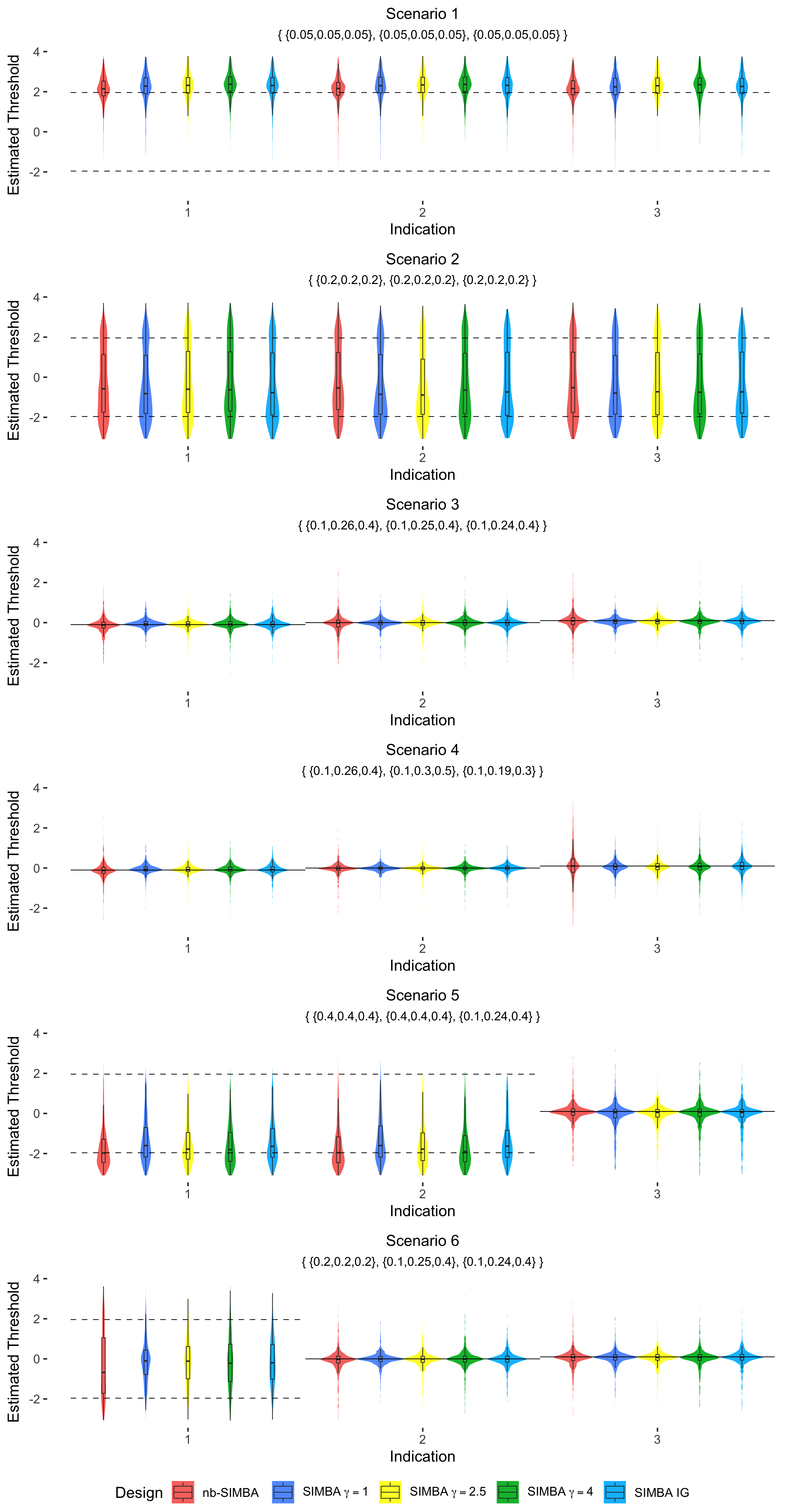} 
    \caption{Estimated Threshold}
    \end{subfigure}
    \caption{Comparison of operating characteristics of nb-SIMBA
       versus  SIMBA under three different
       hyperparameter settings for  $\gamma$ and
      the IG prior for $\sigma_x^2$.  Simulations are under 
      true thresholds $(x_1,x_2,x_3) = (-0.1, 0, 0.1)$. The first
      column panel (a) presents the percentages
       of repeat simulations 
      choosing the  optimal decision under each scenario for SIMBA and nb-SIMBA. The
      second panel (b) presents violin plots of estimated $\hatt_i$
      for SIMBA and nb-SIMBA. The dashed line indicates the 95\% range
      of simulated $X_{ij}$, while the solid line represents the true
      biomarker threshold.} 
    \label{fig:gamma_compare}
\end{figure}

For each scenario in Section \ref{sec:simsetup}, we generate  1,000
simulated trials  and apply SIMBA   under each of the two new $\gamma$
values for the half-Cauchy prior   and the  IG prior.     
The operating characteristics are summarized in Figure
\ref{fig:gamma_compare} including results for the original   nb-SIMBA
and SIMBA approaches configured with $\gamma = 2.5$.  For clarity, the
figure presents only the percentage of selecting the optimal decision
for each indication, as specified in Table
\ref{tab:sc_optimal_decision}.  
When $\gamma = 1$, which enforces stronger borrowing, SIMBA exhibits
higher percentages of selecting the optimal decision $RP$ for the
three indications in scenario 3. However, in scenario 6, where
indication 1 shows no difference,  SIMBA produces  a lower percentage
of selecting the optimal decision $INC$ for indication 1. 

SIMBA exhibits robust estimation of $x_i$ using  $\hatt_i$ under all
three $\gamma$ values for the half-Cauchy prior.  When subgroups are
present for  an indication,  SIMBA exhibits a higher  accuracy in
estimating $x_i$   compared to both nb-SIMBA and SIMBA with the IG
prior. 

Overall,  the  simulation results indicate that SIMBA is a robust method that performs relatively well with information borrowing across indications. 
In addition, SIMBA performs better under the half-Cauchy prior with a $\gamma$ value between 1 and 4.  

\section{ A Basket Trial for Cancer Immunotherapy } \label{sec:example}
We apply SIMBA  to analyze  data from a basket trial
investigating a novel 
therapeutic agent targeting a specific immune marker across three arms
corresponding to three oncology indications.  The first arm 
enrolls 
patients with gastric cancer or gastrointestinal junction (G/GEJ)
cancer $(N_1 = 65)$. The second arm  enrolls  patients with
pancreatic cancer $(N_2 = 35)$. The third arm includes a small group
of patients $(N_3 = 5)$ with various other indications—such as gallbladder cancer, intrahepatic cholangiocarcinoma, extrahepatic
cholangiocarcinoma, and bile duct cancer. Although the third arm is
included in the analysis, results should be interpreted with caution
due to the limited sample size.

\begin{figure}[!htb]
    \centering
    \includegraphics[width=0.6\linewidth]{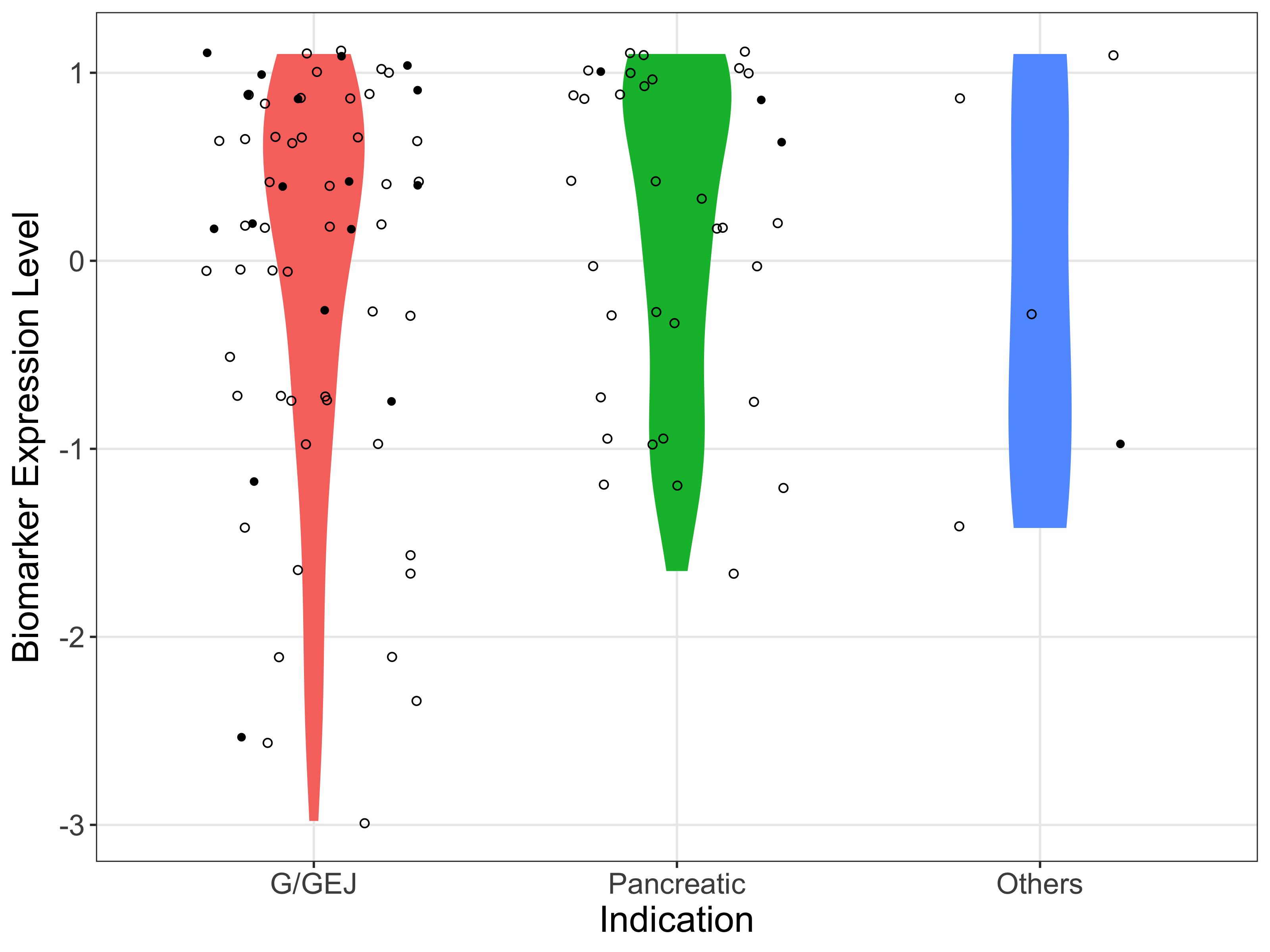}
    \caption{Violin plot of biomarker expression levels $(X_{ij})$
      across arms. Each violin represents the distribution of
      biomarker expression levels for a specific
      indication. Individual patient observations are overlaid as
      points: solid circles denote patients who experienced an
      objective response $(Y_{ij} = 1)$, while hollow circles
      represent non-responders $(Y_{ij} =
      0)$.} \label{fig:example_data} 
\end{figure}

Each patient has an associated biomarker expression level $(X_{ij})$ and a binary response outcome $(Y_{ij})$ indicating whether the patient experienced a clinical response, as defined by objective response rate (ORR) according to the RECIST \citep{eisenhauer2009new} criteria. A summary of the patient-level data used in the analysis is presented in Figure \ref{fig:example_data}. 
Overall, most patients exhibit biomarker expression levels between -1 and 1, with distributions skewed slightly toward 1 in the G/GEJ and pancreatic cancer arms. Solid circles, indicating responders $(Y_{ij} = 1)$, tend to cluster around moderate-to-high expression levels, while hollow circles (non-responders, $Y_{ij} = 0$) are more widely spread and often appear at lower expression levels. This visual trend suggests a potential association between higher biomarker expression and clinical response, particularly in the G/GEJ and pancreatic arms.

\begin{figure}[!htb]
    \centering
    \includegraphics[width=\linewidth]{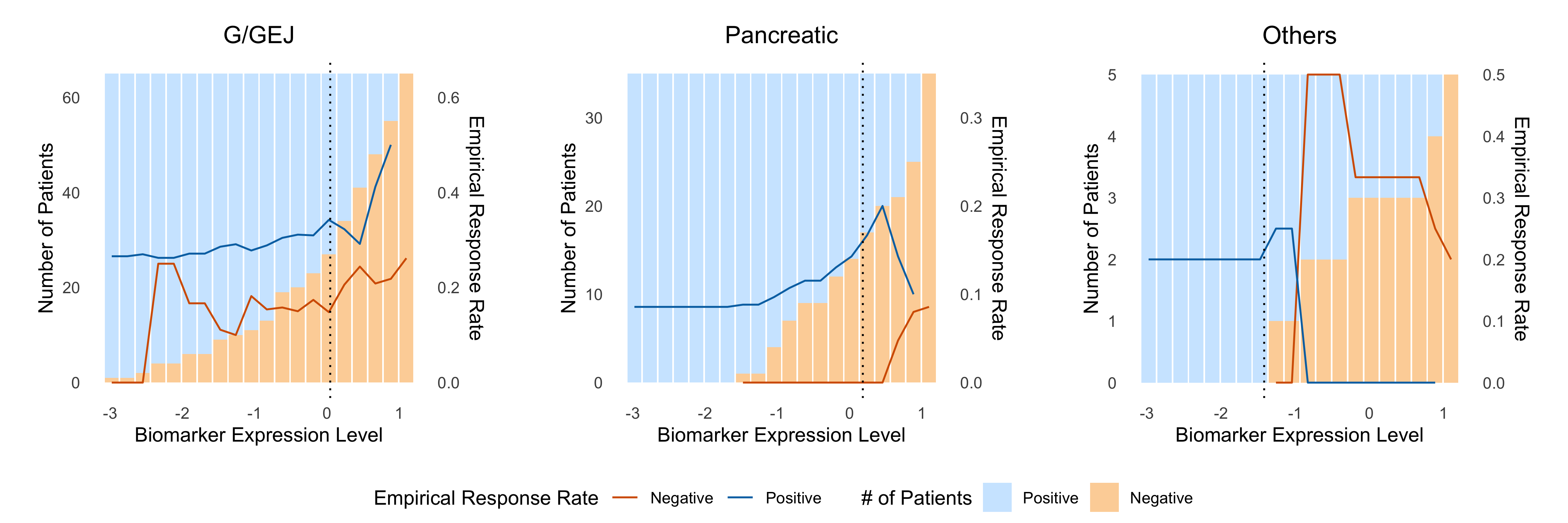}
    \caption{Distribution of patients and empirical response rates
      across thresholds of biomarker expression levels.
       For each biomarker level $x$ the vertical bar shows the
      percentage of patients with biomarkers above $x$, $X_{ij}>x$
      (``positive'') and below (``negative'').   
The lines show the empirical response rates  for both  subgroups,
mapped to the right y-axis: the blue line represents the empirical
response rate among patients in the positive subgroup, and the orange
line represents that of the negative subgroup. The vertical dotted
line indicates the estimated optimal threshold $\hat{t}_i$ of
biomarker expression  under  loss function
\eqref{eq:loss_t}.} \label{fig:example_res} 
\end{figure}

We  set $w_1 = 0.2 $ and $w_2 = 0.5$ in the loss function
\eqref{eq:loss_t} and
 otherwise  follow the parameter settings in Section \ref{sec:simsetup}. 
Figure \ref{fig:example_res} 
shows SIMBA analysis results in terms of biomarker subgroups and their
corresponding response rates for the three indications. For each fixed
biomarker expression level as the horizontal axis, the two vertical
bars represent the numbers of patients with biomarker measurements
greater than (positive subgroup) or less than (negative subgroup) the
corresponding level. The two colored curves track the empirical
response rates of the two subgroups.  
At the estimated optimal threshold $\hat{t}_i$'s (indicated by the vertical dotted line), patients in the positive subgroup consistently show a substantially higher empirical response rate than those in the negative subgroup  for G/GEJ and  pancreatic cancers.   Importantly, the number of patients in the two subgroups remains reasonably balanced at the estimated threshold in both arms, supporting the validity of the threshold as a meaningful discriminator of treatment response. 
These findings suggest that biomarker expression level may serve as a useful stratification variable in identifying likely responders across multiple indications.

In contrast, for the ``Others" arm, all patients fall into the positive subgroup under the estimated threshold, making it impossible to compare response rates between subgroups. The interpretation should be made with caution due to the very limited sample size. As such, the results for this arm are considered exploratory.

Following the decision framework in Section \ref{sec:dec_frame}, we obtain the values of $LRV$ and $TV$ for the three arms from our clinical investigators as follows. For the G/GEJ arm, \(LRV_1 = 0.15\), \(TV_1 = 0.25\); for the pancreatic arm, \(LRV_2 = 0.03\), \(TV_2 = 0.06\); for the ``Others" arm, \(LRV_3 = 0.03\), \(TV_3 = 0.06\). Corresponding interval lengths are set as $\epsilon_1 = 0.05$, $\epsilon_2 = 0.03$, and $\epsilon_3 = 0.03$, respectively. These values reflect differences in clinical expectations and development thresholds across indications, and allow the decision framework to adapt to indication-specific standards of efficacy. 
SIMBA found the optimal final decisions  for the three arms are $A_1^\star = A_2^\star = RA$ while $A_3^\star = RP$  using   the loss function \eqref{eq:loss} and the decision Table \ref{tab:map_tab}. Even though SIMBA suggests the presence of a positive and negative subgroup for the G/GEJ and pancreatic cancers, their response rates exceed the LRV values for both indications and therefore the SIMBA optimal decisions are RA, ``recommend the drug for all patients", for
 both indications.   
As previously noted, the interpretation of decision $RP$ in the
``Others" arm should be made with caution due to the very limited
sample size.

\section{Discussion} \label{sec:discussion}
The   proposed SIMBA method    estimates a  biomarker threshold and
OBS for each indication in a basket trial.   By leveraging a Bayesian
hierarchical model, SIMBA  borrows information across different
indications, thereby improving the precision of threshold estimation.
 We found that the half-Cauchy prior in \eqref{eq:prior_x_m2}
offers a good balance of 
regularization versus  allowing flexibility in
parameter estimates. It stabilizes estimates in non-identifiable or
separated data and enhances predictive performance by balancing large
coefficients with reasonable shrinkage \citep{gelman2008weakly}.  

SIMBA   builds on some of  the  ideas   from \cite{xu2020asied} and
\cite{guo2017bayesian} to form an interval (LRV, TV) for each
indication, and associate trial decisions with the induced efficacy
sub-intervals through a loss  function \eqref{eq:loss} and a mapping function $g(\cdot)$. 

The determination of LRV and TV should be carefully carried out through thorough discussion with the clinical team to align developmental goals and the specific characteristics of the therapeutic area being studied. 

An alternative implementation of SIMBA can be realized
 as the following 2-step procedure: 
first estimate the optimal biomarker thresholds $\hat{t}_i$'s using
the proposed loss function \eqref{eq:loss_t}, and then rerun the MCMC
algorithm conditional on the estimated thresholds to obtain updated
posterior distributions and optimal decisions. This procedure allows
the decision-making step to  more accurately reflect the finalized
biomarker-defined subgroups, and can be particularly beneficial when
the posterior of the threshold is highly concentrated or when clarity
in subgroup definition is desired prior to final
decision-making. Simulation results demonstrating the performance of
this two-step approach are provided in Figure
\ref{fig:op_2s}. However, the two-step app ignores the variablity of
the estimated $\hat{t}_i$.

\section*{Data Availability Statement}
The data and R code used for the application of the proposed method are available in the Supporting Information.

\section*{Acknowledgments}
Peter Müller has been partially funded by NIH/R01MH128085. We used AI-based tools to assist with language polishing and editorial refinement of the manuscript.

\clearpage\newpage
\bibliographystyle{apalike}
\bibliography{seabet}

\clearpage\newpage
\begin{appendices}
\appendixpage
\setcounter{table}{0}
\renewcommand{\thetable}{A.\arabic{table}}
\setcounter{equation}{0}
\renewcommand{\theequation}{A.\arabic{equation}}
\setcounter{figure}{0}
\renewcommand{\thefigure}{A.\arabic{figure}}

\section{Prior Setting}
\begin{figure}[!htbp]
  \centering
  \includegraphics[width=0.8\textwidth]{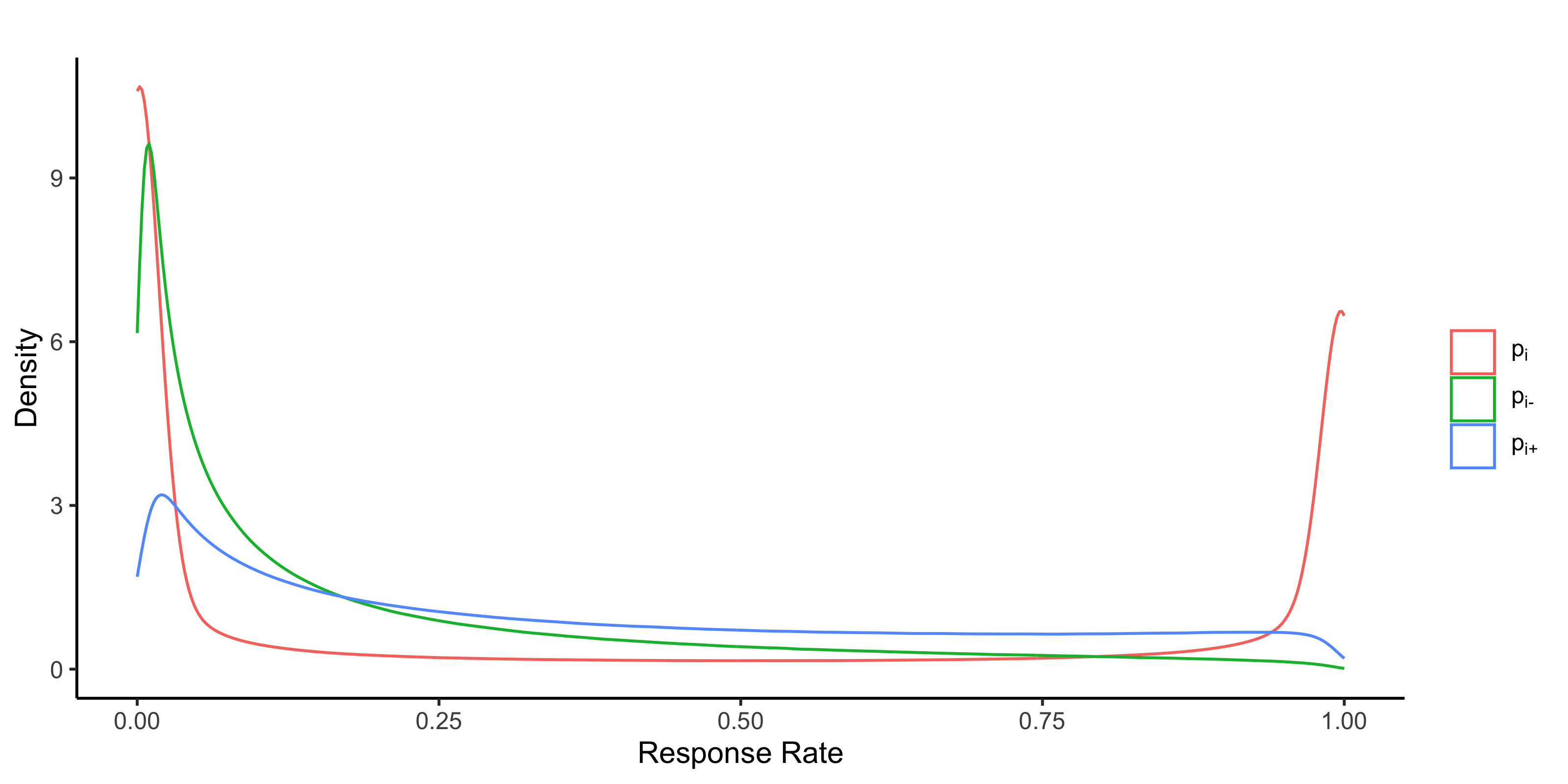}
  \caption{Prior densities of $p_i$, $p_{i-}$, and $p_{i+}$.}
  \label{fig:prior_p}
\end{figure}

\begin{figure}[!htbp]
  \centering
  \includegraphics[width=0.8\textwidth]{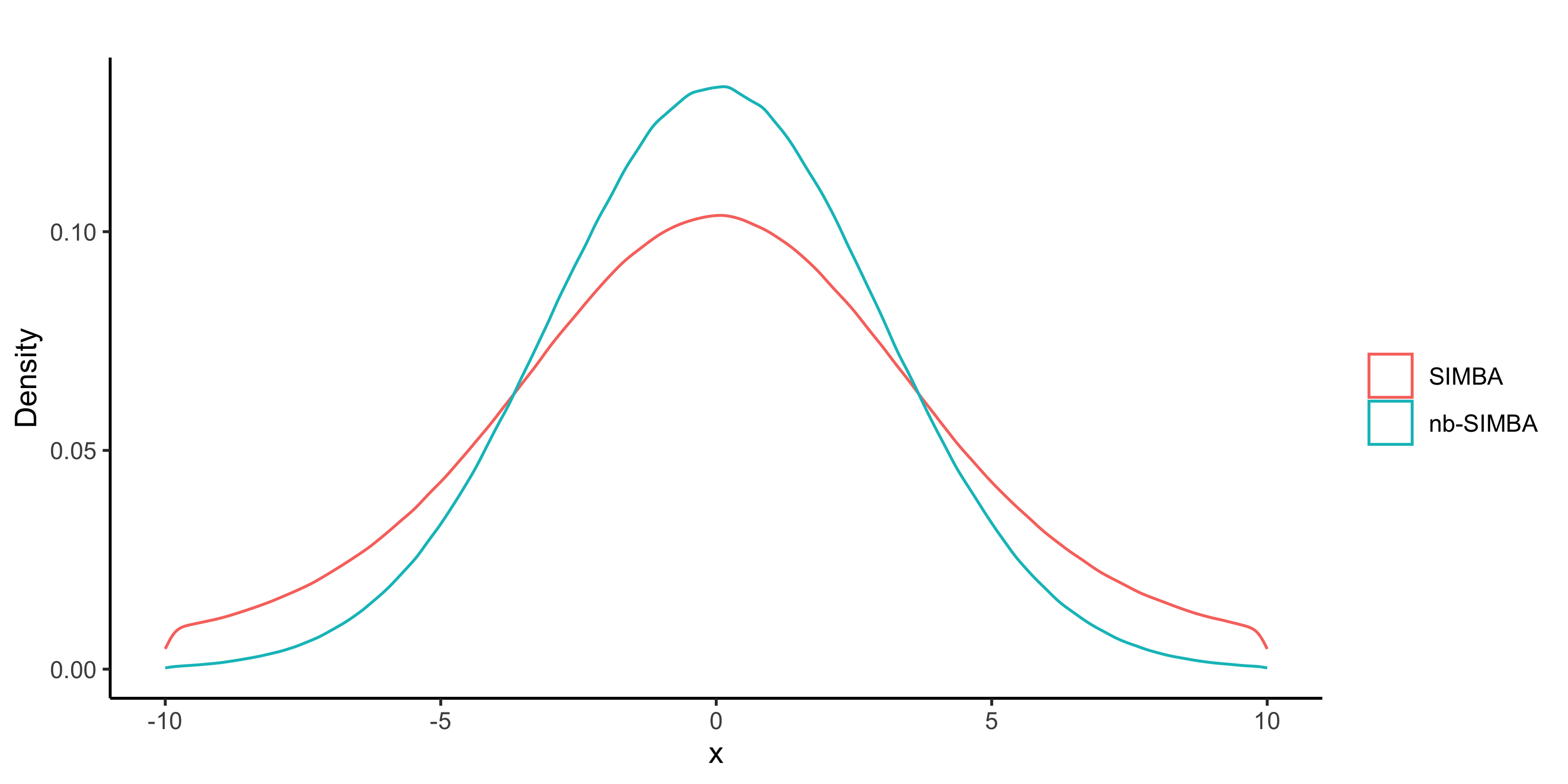}
  \caption{Prior density of $x_i$, $\pi(x_i \mid \mu_x = 0, \sigma_x =3)$, in nb-SIMBA and marginal prior density of $x_i$, $\pi(x_i) = \iint \pi(x_i \mid \mu_x, \sigma_x)\pi(\mu_x)\pi(\sigma_x) d\mu_xd\sigma_x$, in SIMBA.}
  \label{fig:prior_x}
\end{figure}

\begin{figure}[!htbp]
  \centering
  \includegraphics[width=\textwidth]{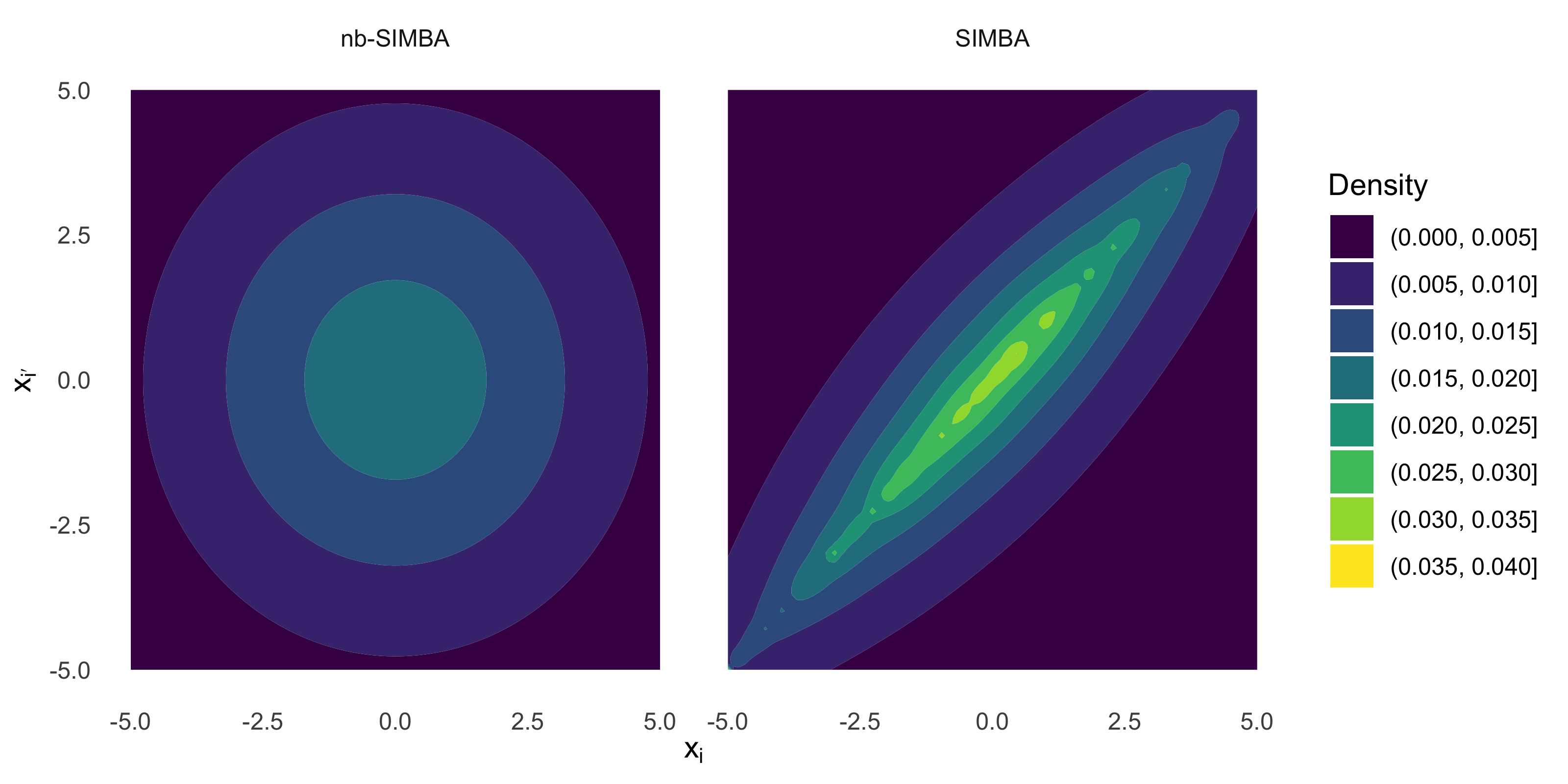}
  \caption{ Conditional joint  prior density of $x_i$ and $x_{i'}$, $\pi(x_i , x_{i'}\mid \mu_x = 0, \sigma_x =3)$, in nb-SIMBA and  conditional joint   prior density of $x_i$ and $x_{i'}$, $\pi(x_i ,  x_{i'}  ) = \iint \pi(x_i , x_{i'}\mid \mu_x, \sigma_x)\pi(\mu_x)\pi(\sigma_x) d\mu_xd\sigma_x$, in SIMBA, where $i \neq i' \in \{1,\cdots,I\}$.}
  \label{fig:joint_prior_x}
\end{figure}
\clearpage
\newpage

\section{Additional Simulation Results}

\subsection{Alternative True Expression Thresholds}
\begin{table}[htbp]
  \centering
  \caption{True response rates of  the positive subgroup, the negative subgroup,   and all patients   in the three indications with their corresponding final and optimal decisions across the six scenarios given the true thresholds of expression levels $(x_1,x_2,x_3) = (0, 0, 0)$.}
    \begin{tabular}{c|ccc|ccc}
    \hline
    \multirow{2}[1]{*}{Scenario} & \multicolumn{3}{c|}{\{$p_{i-},p_{i},p_{i+}$\}}& \multicolumn{3}{c}{$A_i^\star$} \\
\cline{2-7}          & Indication 1 & Indication 2 & Indication 3& Indication 1 & Indication 2 & Indication 3 \\
    \hline
    1     & \{0.05, 0.05, 0.05\} & \{0.05, 0.05, 0.05\} & \{0.05, 0.05, 0.05\} & $S$ & $S$ & $S$\\
    2     & \{0.2, 0.2, 0.2\} & \{0.2, 0.2, 0.2\} & \{0.2, 0.2, 0.2\} & $INC$ & $INC$ & $INC$\\
    3     & \{0.1, 0.25, 0.4\} & \{0.1, 0.25, 0.4\} & \{0.1, 0.25, 0.4\} & $RP$ & $RP$ & $RP$\\
    4     & \{0.1, 0.25, 0.4\} & \{0.1, 0.3, 0.5\} & \{0.1, 0.2, 0.3\} & $RP$ & $RP$ & $INC$\\
    5     & \{0.4, 0.4, 0.4\} & \{0.4, 0.4, 0.4\} & \{0.1, 0.25, 0.4\} & $RA$ & $RA$ & $RP$\\
    6     & \{0.2, 0.2, 0.2\} & \{0.1, 0.25, 0.4\} & \{0.1, 0.25, 0.4\} & $INC$ & $RP$ & $RP$\\
    \hline
    \end{tabular}
  \label{tab:sc_optimal_decision_0}
\end{table}

\begin{figure}[!htb]
    \centering
    \begin{subfigure}{.26\textwidth}
    \centering
    \includegraphics[width=\linewidth]{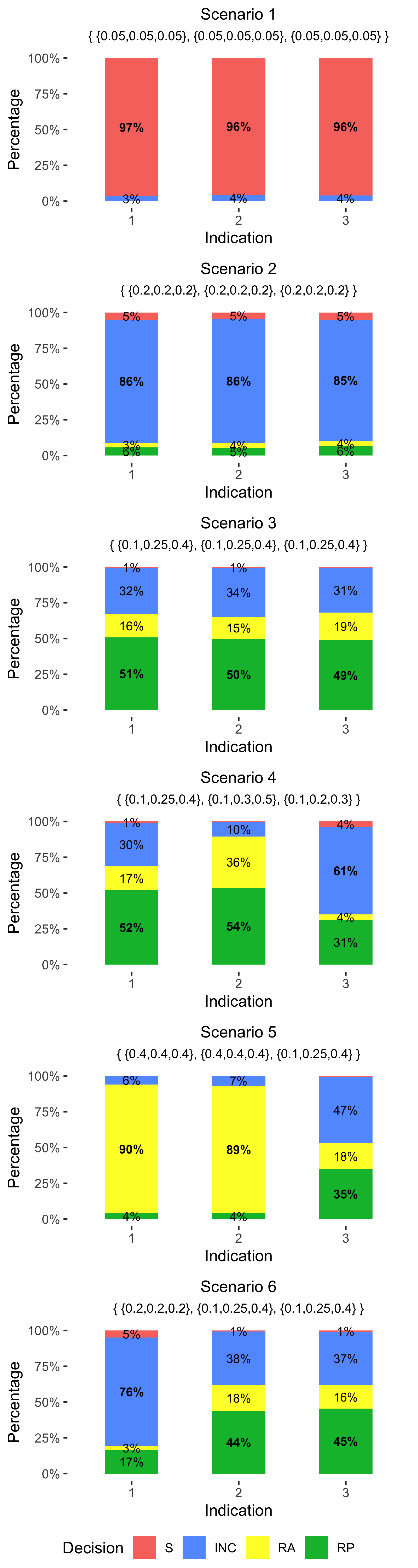}
    \caption{SIMBA}
    \end{subfigure}
    \begin{subfigure}{.26\textwidth}
    \centering
    \includegraphics[width=\linewidth]{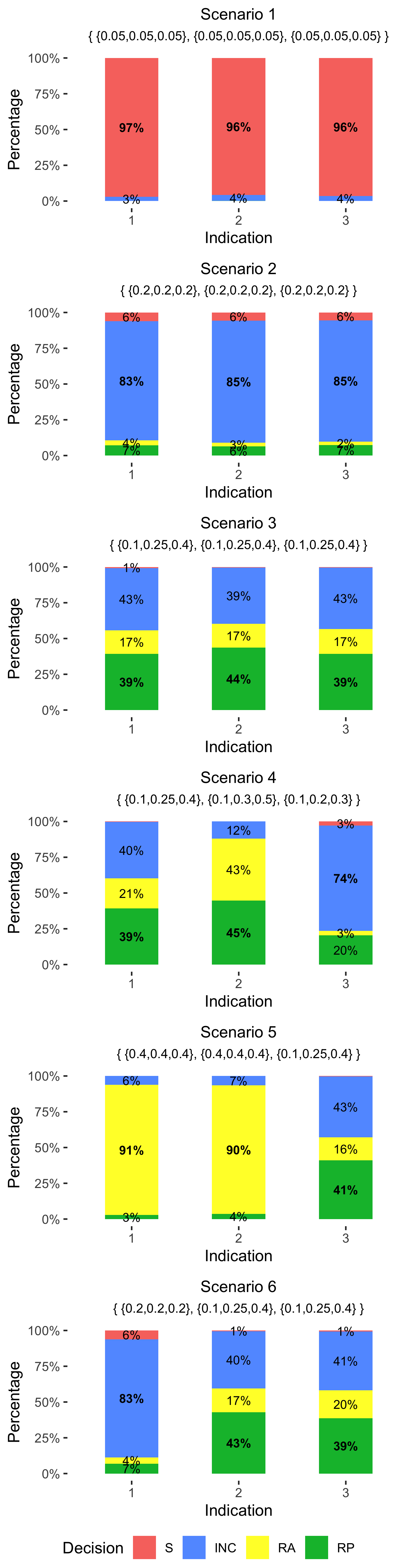} 
    \caption{nb-SIMBA}
    \end{subfigure}
    \begin{subfigure}{.39\textwidth}
    \centering
    \includegraphics[width=\linewidth]{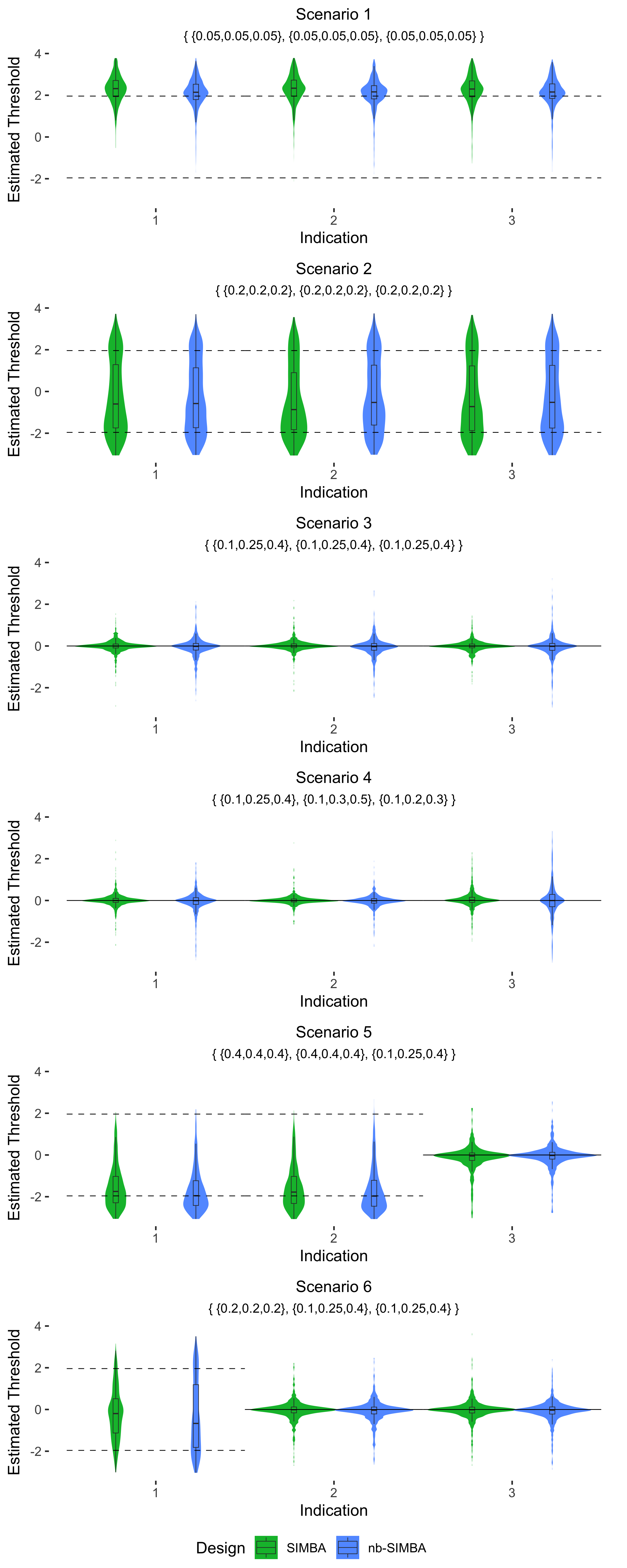} 
    \caption{Estimated Threshold}
    \end{subfigure}
    \caption{Operating characteristics with the true thresholds of expression levels $(x_1,x_2,x_3) = (0, 0, 0)$, where prevalence rates of the three indications are 50\%, 50\%, and 50\%, respectively. The bold font denotes the optimal decision of each indication in each scenario, shown in Table \ref{tab:sc_optimal_decision_0}.
    }
    \label{fig:op_0}
\end{figure}

\begin{table}[htbp]
  \centering
  \caption{True response rates of  the positive subgroup, the negative subgroup,   and all patients    in the three indications with their corresponding final and optimal decisions across the six scenarios given the true thresholds of expression levels $(x_1,x_2,x_3) = (-0.5, 0, 0.5)$.}
    \begin{tabular}{c|ccc|ccc}
    \hline
    \multirow{2}[1]{*}{Scenario} & \multicolumn{3}{c|}{\{$p_{i-},p_{i},p_{i+}$\}}& \multicolumn{3}{c}{$A_i^\star$} \\
\cline{2-7}          & Indication 1 & Indication 2 & Indication 3& Indication 1 & Indication 2 & Indication 3 \\
    \hline
    1     & \{0.05, 0.05, 0.05\} & \{0.05, 0.05, 0.05\} & \{0.05, 0.05, 0.05\} & $S$ & $S$ & $S$\\
    2     & \{0.2, 0.2, 0.2\} & \{0.2, 0.2, 0.2\} & \{0.2, 0.2, 0.2\} & $INC$ & $INC$ & $INC$\\
    3     & \{0.1, 0.31, 0.4\} & \{0.1, 0.25, 0.4\} & \{0.1, 0.19, 0.4\} & $RA$ & $RP$ & $RP$\\
    4     & \{0.1, 0.31, 0.4\} & \{0.1, 0.3, 0.5\} & \{0.1, 0.16, 0.3\} & $RA$ & $RP$ & $INC$\\
    5     & \{0.4, 0.4, 0.4\} & \{0.4, 0.4, 0.4\} & \{0.1, 0.19, 0.4\} & $RA$ & $RA$ & $RP$\\
    6     & \{0.2, 0.2, 0.2\} & \{0.1, 0.25, 0.4\} & \{0.1, 0.19, 0.4\} & $INC$ & $RP$ & $RP$\\
    \hline
    \end{tabular}
  \label{tab:sc_optimal_decision_05}
\end{table}

\begin{figure}[!htb]
    \centering
    \begin{subfigure}{.26\textwidth}
    \centering
    \includegraphics[width=\linewidth]{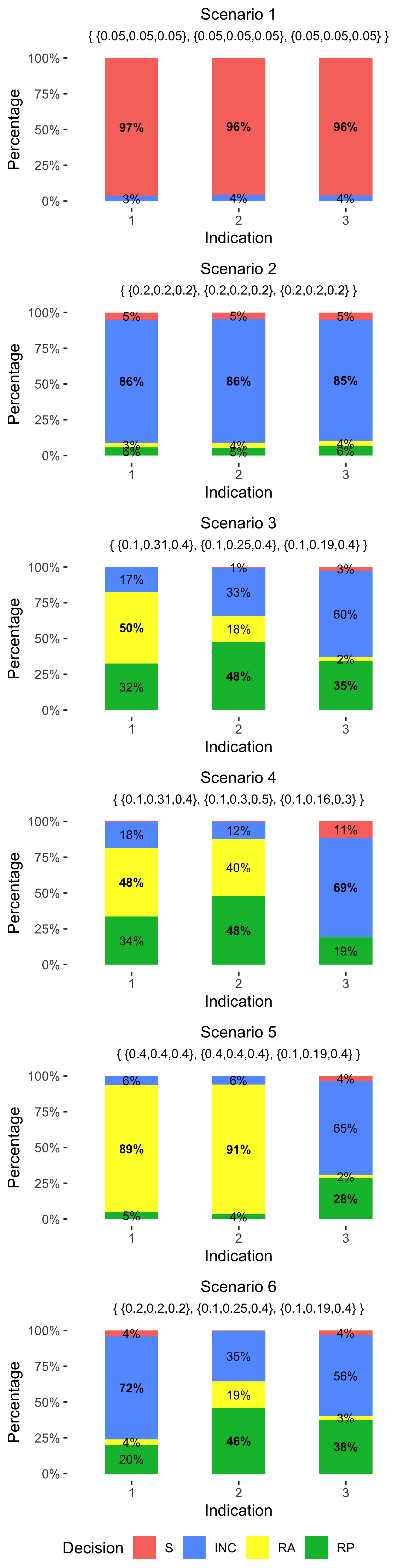}
    \caption{SIMBA}
    \end{subfigure}
    \begin{subfigure}{.26\textwidth}
    \centering
    \includegraphics[width=\linewidth]{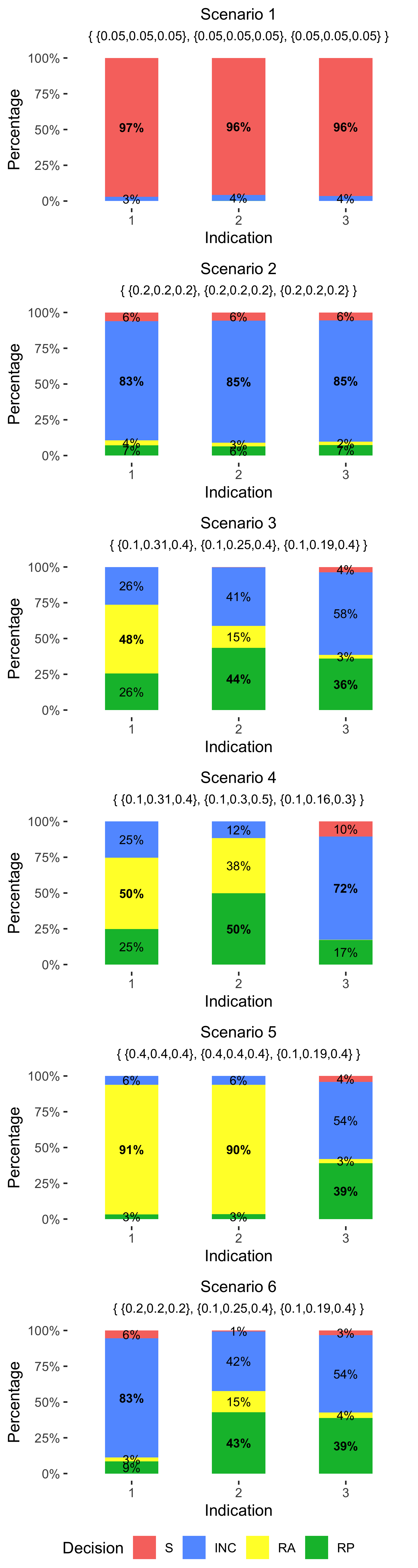} 
    \caption{nb-SIMBA}
    \end{subfigure}
    \begin{subfigure}{.39\textwidth}
    \centering
    \includegraphics[width=\linewidth]{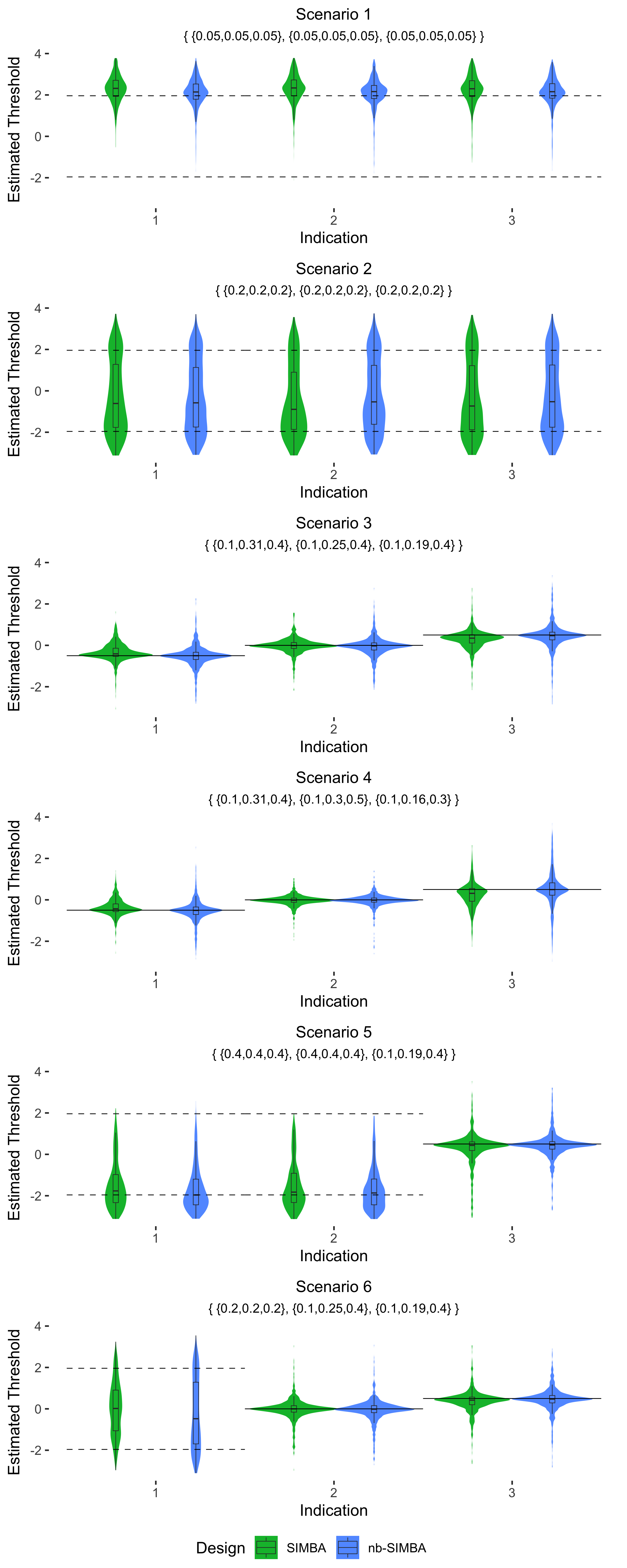} 
    \caption{Estimated Threshold}
    \end{subfigure}
    \caption{Operating characteristics with the true thresholds of expression levels $(x_1,x_2,x_3) = (-0.5, 0, 0.5)$, where prevalence rates of the three indications are around 69\%, 50\%, and 31\%, respectively. The bold font denotes the optimal decision of each indication in each scenario, shown in Table \ref{tab:sc_optimal_decision_05}.
    }
    \label{fig:op_05}
\end{figure}

\clearpage\newpage
\subsection{Alternative Weights}
\begin{figure}[!htb]
    \centering
    \includegraphics[width=0.95\linewidth]{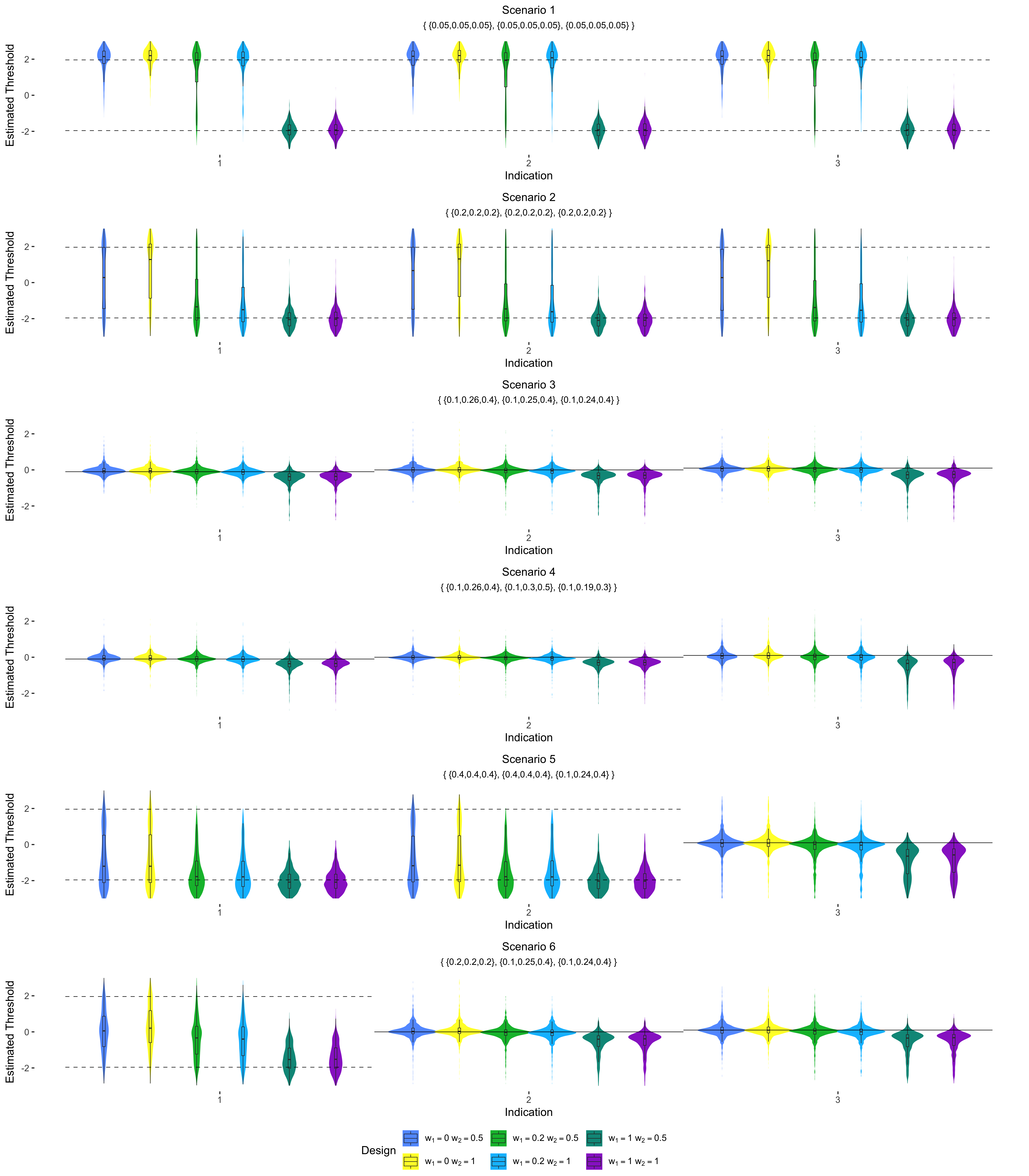}
    \caption{Violin plots of the estimated biomarker thresholds for SIMBA under different weight settings in the loss function \eqref{eq:loss_t}.
    }
    \label{fig:w_compare}
\end{figure}

\clearpage\newpage
\subsection{Alternative Two-step Approach}
\begin{figure}[!htb]
    \centering
    \begin{subfigure}{.26\textwidth}
    \centering
    \includegraphics[width=\linewidth]{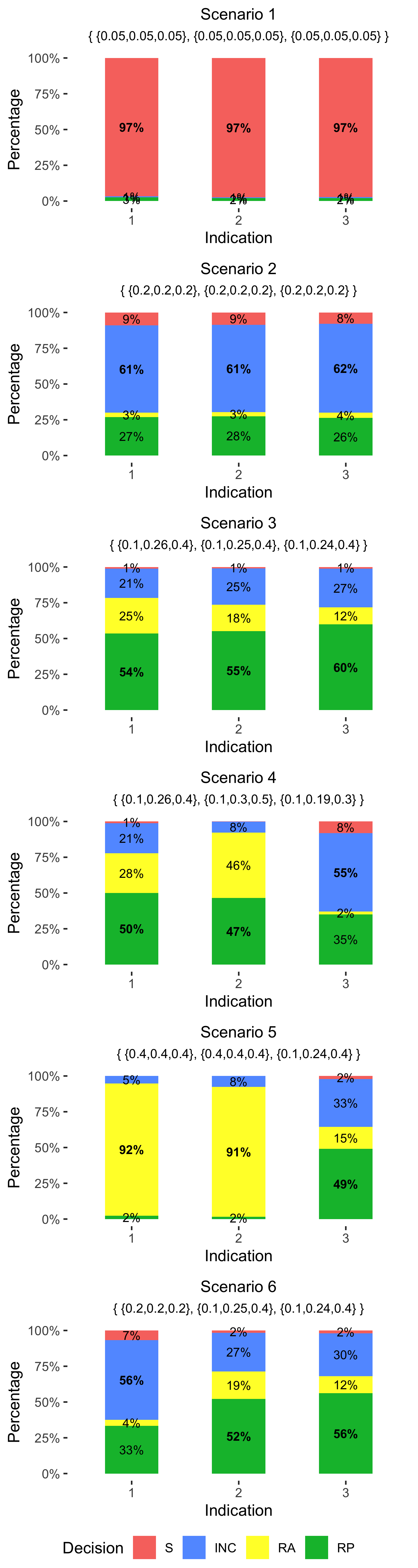}
    \caption{SIMBA}
    \end{subfigure}
    \begin{subfigure}{.26\textwidth}
    \centering
    \includegraphics[width=\linewidth]{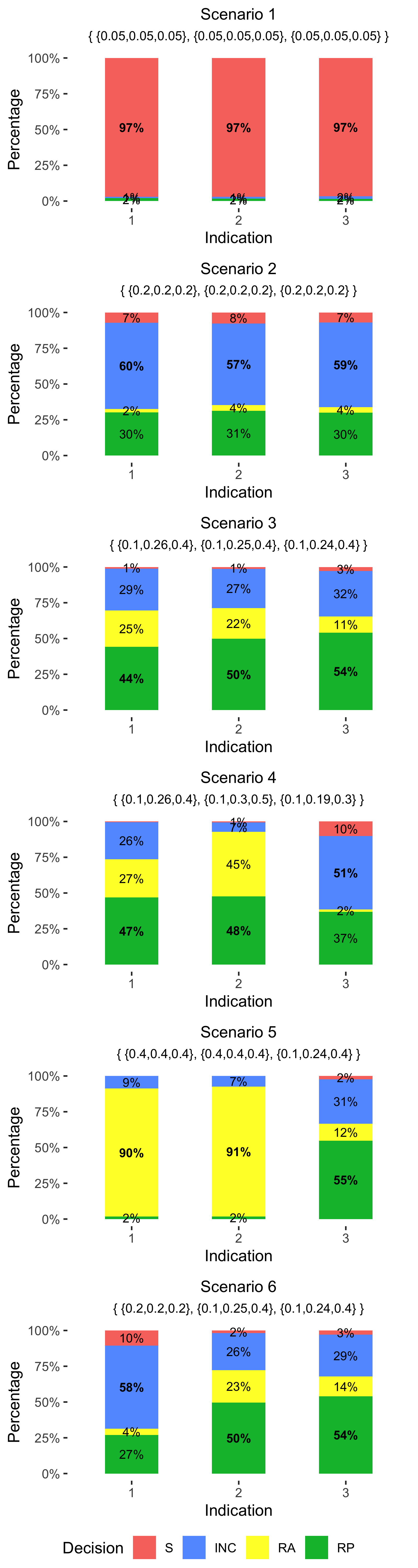} 
    \caption{nb-SIMBA}
    \end{subfigure}
    \caption{Operating characteristics with the true thresholds of expression levels $(x_1,x_2,x_3) = (-0.1, 0, 0.1)$ for two-step SIMBA and nb-SIMBA. The first two column panels (a) and (b)  present  the percentages of four distinct final decisions under each scenario, where the bold font denotes the optimal decision. 
    }
    \label{fig:op_2s}
\end{figure}

\end{appendices}
\end{document}